\documentclass[conference]{IEEEtran}

\usepackage{xurl}
\usepackage{hyperref}
\usepackage{tikz}
\usepackage{amsmath}
\usepackage{algorithmic}            
\usepackage{algorithm}
\usepackage{multirow}
\usepackage{tabularx}
\usepackage{subcaption}
\usepackage[prologue, table, dvipsnames, table, x11names]{xcolor}

\usepackage{enumitem}
\usepackage{caption}
\usepackage{annotate-equations}
\usepackage{float}
\usepackage{makecell}
\usepackage{nicematrix}
\usepackage{soul}
\usepackage{diagbox}
\usepackage{bm}
\usepackage{comment}
\usepackage{svg}

\usetikzlibrary{arrows,shadows, positioning}
\usepackage{pgf-umlsd}

\RenewDocumentEnvironment{messcall}{ O{1} m O{} m m }{
  \stepcounter{seqlevel}
  \stepcounter{callevel} %
  \path
  (#2)+(0,-\theseqlevel*\unitfactor-0.7*\unitfactor) node (cf\thecallevel) {}
  (#5.\threadbias)+(0,-\theseqlevel*\unitfactor-0.7*\unitfactor) node (ct\thecallevel) {};
  
  \draw[->,>=angle 60] ({cf\thecallevel}) -- (ct\thecallevel)
  node[midway, above, #3] {#4};
  \def\l\thecallevel{#1}
  \def\f\thecallevel{#2}
  \def\t\thecallevel{#5}
  \tikzstyle{threadstyle}+=[instcolor#2]
}
{
  \addtocounter{seqlevel}{\l\thecallevel}
  \path
  (\f\thecallevel)+(0,-\theseqlevel*\unitfactor-0.7*\unitfactor) node (rf\thecallevel) {}
  (\t\thecallevel.\threadbias)+(0,-\theseqlevel*\unitfactor-0.3*\unitfactor) node (rt\thecallevel) {};
  \drawthread{ct\thecallevel}{rt\thecallevel}
  \addtocounter{callevel}{-1} %
}

\newcommand\algorithmicprocedure{\textbf{func}}
\newcommand{\algorithmicendprocedure}{\algorithmicend\ \algorithmicprocedure}
\makeatletter
\newcommand\PROCEDURE[3][default]{%
  \ALC@it
  \algorithmicprocedure\ \textsc{#2}(#3)%
  \ALC@com{#1}%
  \begin{ALC@prc}%
}
\newcommand\ENDPROCEDURE{%
  \end{ALC@prc}%
  \ifthenelse{\boolean{ALC@noend}}{}{%
    \ALC@it\algorithmicendprocedure
  }%
}
\newenvironment{ALC@prc}{\begin{ALC@g}}{\end{ALC@g}}
\makeatother

\usepackage{pifont}%

\definecolor{normal}{HTML}{008fd5}
\definecolor{op}{HTML}{0FFF50}
\definecolor{attack}{HTML}{fc4f30}
\definecolor{req}{HTML}{FFC61E}
\definecolor{tbdNormal}{HTML}{ADD8E6}
\definecolor{tbdAttack}{HTML}{FFC0CB}
\definecolor{tbdSysPOV}{HTML}{90EE90}

\tikzset{
    cross/.pic = {
    \draw[rotate = 45] (-#1,0) -- (#1,0);
    \draw[rotate = 45] (0,-#1) -- (0, #1);
    }
}

\newcommand{\TBD}{tBD}
\newcommand{\circleMission}{
    \begin{tikzpicture}
        \filldraw[color=blue!60, fill=white!5, line width=1pt, anchor=south](-1,0) circle (0.15);
    \end{tikzpicture}
}
\newcommand{\linearMission}{
    \begin{tikzpicture}
        \draw[thick, blue] (0,0) node[anchor=south]{}
          -- (0.25,0) node[anchor=south]{}
          -- (0.25,0.25) node[anchor=south]{}
          -- cycle;
    \end{tikzpicture}
}
\newcommand{\holdMission}{
    \begin{tikzpicture}
        \draw[line width=0.5pt, color=blue] plot [domain=-0.1:0.1, samples=32, smooth] ({0.1*rand},{0.1*rand});
    \end{tikzpicture}
}

\newcommand{\circleMissionSmall}{
    \begin{tikzpicture}
        \filldraw[color=blue!60, fill=white!5, line width=1pt, anchor=south](-1,0) circle (0.1);
    \end{tikzpicture}
}
\newcommand{\linearMissionSmall}{
    \resizebox{!}{1em}{
        \begin{tikzpicture}
            \draw[thick, blue] (0,0) node[anchor=south]{}
              -- (0.3,0) node[anchor=south]{}
              -- (0.3,0.3) node[anchor=south]{}
              -- cycle;
        \end{tikzpicture}
    }
}
\newcommand{\holdMissionSmall}{
    \begin{tikzpicture}
        \draw[line width=0.5pt, color=blue] plot [domain=-0.1:0.1, samples=32, smooth] ({0.05*rand},{0.05*rand});
    \end{tikzpicture}
}

\newcommand{\greencheck}{{\color{green} \ding{51}}}%
\newcommand{\redxmark}{{\color{red} \ding{55}}}%

\DeclareRobustCommand\circledSeq[1]{\tikz[baseline=(char.base)]{\node[shape=circle,fill=NavyBlue,draw=black,inner sep=1pt] (char) {\textcolor{white} {#1}};}}

\DeclareRobustCommand\circledSeqSmaller[1]{\tikz[baseline=(char.base)]{\node[shape=circle,fill=NavyBlue,draw=black,inner sep=0.25pt] (char) {\textcolor{white} {#1}};}}

\DeclareRobustCommand\circledReq[1]{\tikz[baseline=(char.base)]{\node[shape=circle,fill=black,draw=black,inner sep=1pt] (char) {\textcolor{white} {\textbf{#1}}};}}

\makeatletter
 \newcommand{\iddef}[1]{\Hy@raisedlink{\hypertarget{#1}{}}#1}
\makeatother
\makeatletter
\newcommand\footnoteref[1]{\protected@xdef\@thefnmark{\ref{#1}}\@footnotemark}
\makeatother

\newcommand{\eq}[1]{Equation~#1}
\newcommand{\fig}[1]{Figure~#1}
\newcommand{\tab}[1]{Table~#1}
\newcommand{\alg}[1]{Algorithm~#1}
\newcommand{\app}[1]{Appendix~#1}
\newcommand{\paperName}{Requiem\xspace}
\newcommand{\papername}{\textsc{Requiem}\xspace}
\newcommand{\paperwebsite}{\texttt{https://projrequiem.github.io}\xspace}

\newcommand{\psection}[1]{\noindent\textbf{#1.}}

\newcommand{\rqu}[1]{\textbf{RQ#1}}
\newcommand{\dqu}[1]{\textbf{DQ#1}}

\newcommand{\wrt}{{\it w.r.t.}\xspace}
\newcommand{\ie}{{\it i.e.,}\xspace}
\newcommand{\eg}{{\it e.g.,}\xspace}
\newcommand{\etc}{{\it etc.}\xspace}

\newcommand{\viz}{\textit{viz.}\xspace}
\newcommand{\ci}{{\it (i)} }
\newcommand{\cii}{{\it (ii)} }
\newcommand{\ciii}{{\it (iii)} }
\newcommand{\civ}{{\it (iv)} }
\newcommand{\cv}{{\it (v)} }
\newcommand{\ca}{{\it (a) }}
\newcommand{\cb}{{\it (b) }}
\newcommand{\cc}{{\it (c) }}

\newcommand{\note}{{\bf Note: }}

\newif\ifdraft
\draftfalse %
\endlinechar=`\^^M

\usepackage{iftex}
\ifPDFTeX
    \usepackage[T1]{fontenc}
    \usepackage[utf8]{inputenc}
\else
   \ifXeTeX
     \usepackage{fontspec}
   \else 
     \usepackage{luatextra}
   \fi
   \defaultfontfeatures{Ligatures=TeX}
\fi

\usepackage{blindtext}
\usepackage{extdash}

\usepackage{sepfootnotes}

\newcommand{\hairspace}{\ifmmode\mskip1mu\else\kern0.08em\fi}

\newfootnotes{f}

\newcommand{\insight}[1]{
    \noindent\fbox{
        \parbox{0.96\columnwidth}{
            \psection{Insight}
            #1
        }
    }
}

\usepackage{cite}
\usepackage{amsmath}
\usepackage{array}
\drafttrue

\begin{document}
\title{Predictable by Design, Vulnerable by Nature: Security Consequences of Learnability in UAV State Estimators}

\author{
    \IEEEauthorblockN{
        Kyo Hyun Kim\IEEEauthorrefmark{1},
        Tejaaswini Narendran\IEEEauthorrefmark{2},
        Bijan Mehralizadeh\IEEEauthorrefmark{2}, 
        Zaid Abu-Abbas\IEEEauthorrefmark{2},\\
        Denizhan Kara\IEEEauthorrefmark{1},
        Vineetha Paruchuri\IEEEauthorrefmark{2},
        Sibin Mohan \IEEEauthorrefmark{2},
        Greg Kimberly \IEEEauthorrefmark{3},\\
        Jae Kim \IEEEauthorrefmark{3}, and
        Josh Eckhardt \IEEEauthorrefmark{3}
    }
    \IEEEauthorblockA{
        \IEEEauthorrefmark{1} UIUC\\
        \IEEEauthorrefmark{2} GWU\\
        \IEEEauthorrefmark{3} Boeing Research and Technology
    }
}

\IEEEoverridecommandlockouts
\makeatletter\def\@IEEEpubidpullup{6.5\baselineskip}\makeatother

\maketitle

\begin{abstract}
    If a mathematical function can be learned from its input/output behavior alone, can an adversary exploit that ``learnability'' to compromise it? 
    What if the function is core to estimating the state of, and controlling, unmanned vehicles such as drones?
    We investigate this question by targeting state estimators in unmanned aerial vehicles (UAVs) --- specifically Extended Kalman Filters (EKFs), an industry standard for autonomous systems, whose inherent modeling of uncertainties and sensor noise create an adversarial space that ML can exploit.

    We present \papername{}, a machine-learning based framework for investigating such vulnerabilities. 
    Our framework functions by, 
    \ci constructing \textit{deep surrogate models} that emulate the state estimation update function using only observed inputs and outputs and
    \cii optimizing \textit{spoofer models} to manipulate sensor values so that they're not easily detectable by standard anomaly detectors --- all of which results in \textit{physical deviations} by the autonomous vehicle.
    We evaluate \papername{}'s efficacy against both, standard PX4 controllers and the state-of-the-art SAVIOR anomaly detector.
    Across \textit{real-world quadrotor experiments} and high-fidelity simulations (Gazebo/PX4), \papername{} demonstrates significant deviations from planned mission paths while evading anomaly detection methods, without the need for intrusive root/administrative access.  
    Our findings suggest that the very properties making state estimators reliable may constitute a security liability, motivating investigation into ``learnability'' as an attack surface in safety-critical systems.

\end{abstract}


\section{Introduction}
\label{sec:introduction}

\fnotecontent{foot:observationalerror}{%
Observational Error (also referred to as Measurement Error) is the difference between the empirically observed (via measurement) value of a parameter and its unknown ``true'' value (unknown, because the exact ``true'' value can only be known via a measurement process that is provably free of error, but such a measurement process does not exist in practice).
}

Recent events~\cite{news2025germanydronesighting, news_2025_ukraine_drone} highlight the growing use of Unmanned Aerial Vehicles (UAVs) in military and civilian applications.
{From an opensource flight controller finding its way into active conflict~\cite{news2025ukrainewarardupilot} to growing interest in counter UAV capabilities~\cite{news_2025_nato_counter_drone},
the security of autonomous vehicles (AVs) is becoming a concern --- not only in military settings but also in civilian domains such as agriculture and disaster response \cite{lee2020deep}.
}
An increasing proportion of safety critical operations that once required manual intervention/control are starting to be delegated to algorithms.
UAVs depend on sensors and state estimation algorithms for correct operation, especially in the autonomous realm.  
{This raises the following question: can systems whose mathematical foundations were designed solely for estimation accuracy provide adversaries with opportunities for malicious actions?
}

State estimation algorithms (\eg Extended Kalman Filter~\cite{kalman1960new}) model the state\footnote{State means different things in different contexts, but we can think of it as a description of the \textit{physical system} using a set of parameters (called ``state variables''). State in itself is static, \ie it does not offer any information about how/whether a system evolves over time. State estimation algorithms \textit{modify} the state.} of the system while accounting for factors such as obstacles, even environmental factors (\eg wind shear) or even failures (engine failures, sensor failures).
All of these, along with the fact that such algorithms attempt to linearize an inherently non-linear system (the real, physical world is hard to model using digital systems) \textit{introduce measurement errors and/or noise}. 
Even the most meticulously designed control systems \textit{always} operate with inherent error due to unavoidable 
sensor measurement inaccuracies and the complexity of modeling vehicular dynamics. 
AVs rely on sensors to perceive their environments and deviations from the expected operational parameters, {\em inaccurate observations can prevent any corrections}.

\begin{figure}[htb]
    \centering
    \begin{minipage}{\columnwidth}
            \def\legendPadding{\hspace{2mm}}
            \footnotesize
            \centering
            \begin{tabular}{|@{\legendPadding}l@{\legendPadding}l@{\legendPadding}l@{\legendPadding}l@{\legendPadding}|}\hline
                \begin{tikzpicture}[scale=1]
                    \draw[line width=0.5mm, color=normal] (0,0) -- (0.5,0);
                \end{tikzpicture}
                : Nominal 
                &
                \begin{tikzpicture}[scale=1]
                    \draw[line width=0.5mm, color=normal, dashed] (0,0) -- (0.5,0);
                \end{tikzpicture}
                : Error Bounds
                &
                \begin{tikzpicture}[scale=1]
                    \draw[line width=0.5mm, color=attack] (0,0) -- (0.5,0);
                \end{tikzpicture}
                : Attack
                &
                \begin{tikzpicture}[scale=1]
                    \draw[line width=1.5mm, color=lightgray] (0,0.1) -- (0.5,0.1);
                \end{tikzpicture}
                : Difference
                \\\hline
            \end{tabular}
        \end{minipage}
        \begin{minipage}{\columnwidth}
            \begin{tikzpicture}
            \begin{scope}[xshift=0cm]
                \node[anchor=south west,inner sep=0] (image) at (0,0) {\includegraphics[width=\columnwidth,height=1.5in]{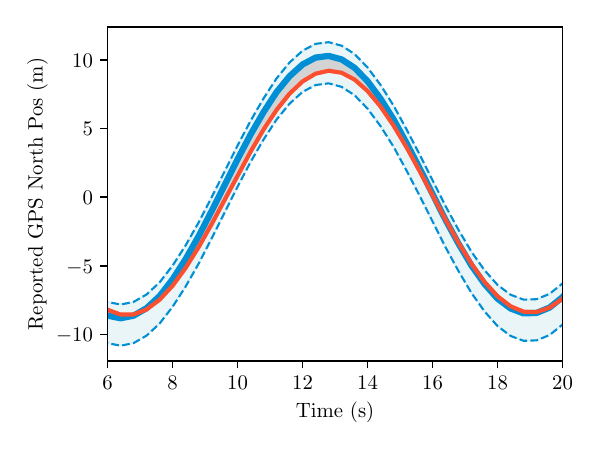}};
                \begin{scope}[x={(image.south east)},y={(image.north west)}]
                    \draw[black, thick, dashed, rotate around={-50:(0.25,0.2)}] (0.18,0.4) rectangle (0.25,1.15);
                    \draw[black, thick, dashed, rotate around={45:(0.83,0.225)}] (0.83,0.225) rectangle (0.9,0.85);
                \end{scope}
                \node[text width=3.25cm, anchor=south west, rotate=40] at (3,0.75) {\footnotesize Accumulating the error};
                \node[text width=2.25cm, anchor=south west, rotate=-45] at (6.2,3.0) {\footnotesize\centering Blend back to nominal};
            \end{scope}
            \end{tikzpicture}%
        \end{minipage}
    \vspace{-2mm}
    \caption{
    Problem space exploitation.
    Nominal mission trace (blue) vs same mission under stealthy attack (red).
    Grey region is the difference being exploited (\ie accumulated error).
    By the end, the vehicle's true position differs from its reported GNSS position.
    Note that the reported GNSS position is \textit{similar} to the nominal.
    }
    \label{fig:example_exploited_space}
    \vspace{-1em}
\end{figure}
This \textit{creates an exploitable space for adversaries}, at times without the risk of detection
(\ie a problem space \cite{pierazzi2020intriguing}).
Figure \ref{fig:example_exploited_space}
provides an example,
if the ``attack values'' (aka spoofed sensor values) lie within the nominal region (as shown by the solid red line), then the state estimation algorithms as well as the anomaly detectors are \textit{unable to detect the false values}.

\begin{figure}[htb]
    \centering
    \begin{minipage}{\columnwidth}
        \begin{minipage}{0.50\linewidth}
            \includegraphics[width=\linewidth]{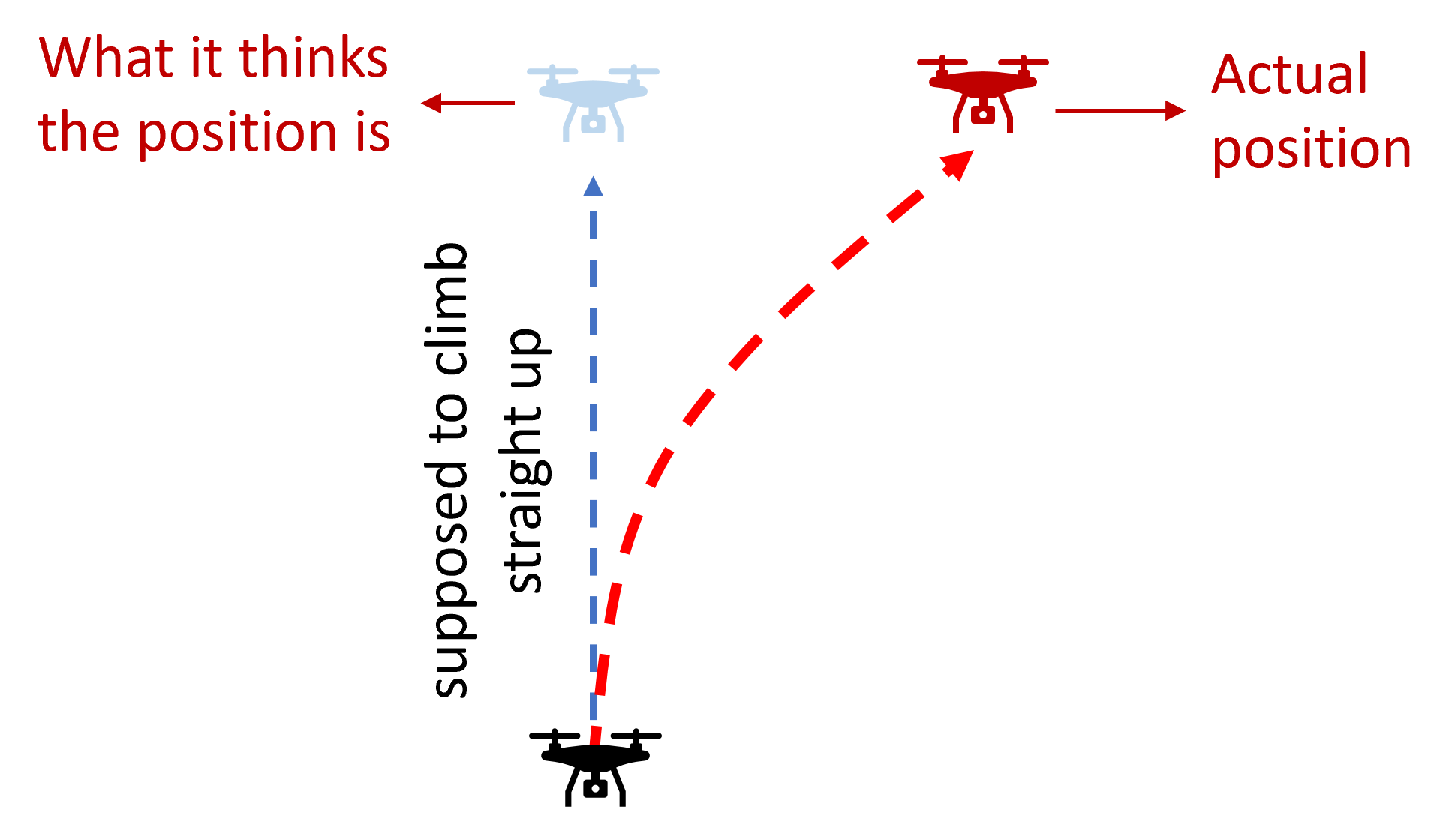}
            \footnotesize
            \begin{tabular}{cc}
                \begin{tikzpicture}[scale=1]
                    \draw[line width=0.3mm, blue] (0,0) -- (0.75,0);
                \end{tikzpicture}
                : Nominal 
                    & 
                \begin{tikzpicture}[scale=1]
                    \draw[line width=0.3mm, red] (0,0) -- (0.75,0);
                \end{tikzpicture}
                : Attack
                \\
            \end{tabular}
        \end{minipage}
        \hspace{2mm}
        \begin{minipage}{0.25\linewidth}
            \includegraphics[width=\linewidth]{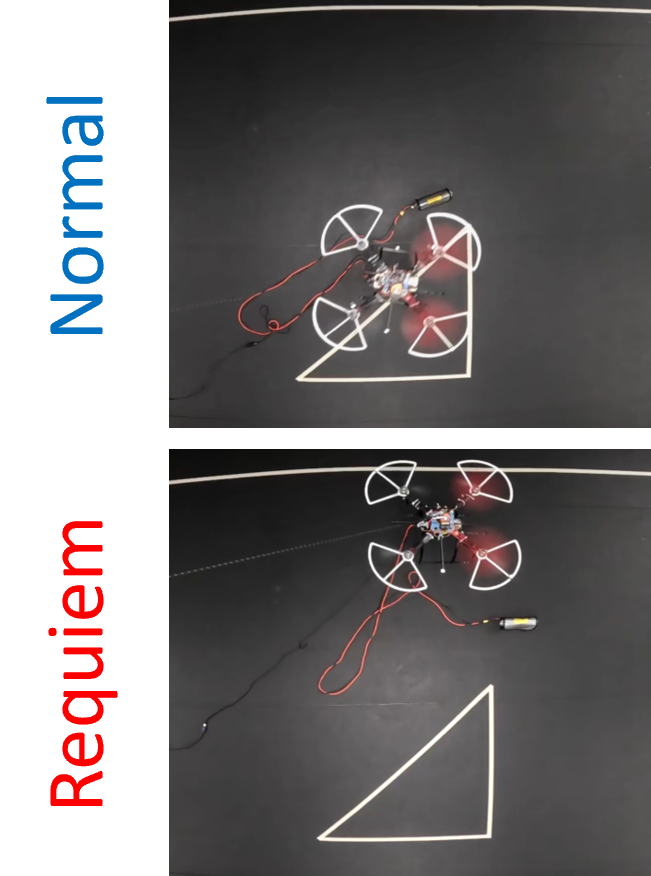}
        \end{minipage}
        \centering
    \end{minipage}
    \caption{\label{fig:example_sa} Example of stealthy attack: the vehicle thinks it is following the mission path (blue) while in actuality, it is deviating (red). The right figure shows the application of \papername{} on a real drone (see \S\ref{sec:rq:realworld}); screenshots of a nominal flight (top) and \papername{} attack (bottom) during a triangle mission. A right triangle with two 1m sides is taped on the floor for scale.
    }
    \vspace{-1em}
\end{figure}

\papername{}, our general framework, \textit{exploits} this very space by generating such spoofed sensor input values --- the end result being that drones \textit{physically deviate} away from their mission objectives.
\fig{\ref{fig:example_sa}} shows examples of real UAVs \footnote{We use UAVs and ``drones'' interchangeably in this paper.} deviating away from expected missions while not realizing that disruptions have occurred!
Although the algorithms (or anomaly detectors) on the UAV should have detected these deviations and corrected for them,
the reported sensor values during the attack remain \textit{very close} to the nominal/expected values (as shown earlier in \fig{\ref{fig:example_exploited_space}}).

Existence of such problem spaces in state estimation have been studied \cite{stealthy_attack_resilient_linear} and shown to exist for EKF, the de-facto standard for vehicles with non-linear dynamics \cite{stealthy_optimal_myopic}.
Prior work has demonstrated this through numerical simulations \cite{arnstrom2024stealthy}, theoretical models \cite{bouabdallah2007full}~\cite{khazraei2023vulnerability} and simulations with simpler vehicle dynamics (\ie cars) \cite{{khazraei2022learning}}.
However, \textit{vehicular dynamics for a UAV are more complex since they must constantly balance in the air}.
Other work attacks the perception module on the PX4\footnote{PX4 is a very popular opensource flight controller used in various types of vehicles} system \cite{amir_stealthy_perception_resiliency,amir_stealthy_perception}, but these approaches \textit{require full internal knowledge} of the state estimator.
Similarly, ARES \cite{ding2023get} uses reinforcement learning based approaches to find adversarial values for a control task to move the vehicle away from its planned path but also requires knowledge of implementation logic and its objective isn't avoiding detection.
Stealth is critical for adversaries, especially in long lasting missions, since early detection allows operators to correct the issue or deploy alternate drones, thwarting the disruption.
Much work has gone into the use of physical spoofing attacks (\eg \cite{kerns2014unmanned, tractorbeam, stealthy_gps_navigation, fusionripper, attack_vec_imu_sound, walnut, poltergeist, kim2024systematic, davidson2016controlling})
but all of them either require line of sight conditions and/or the use of specialized hardware, both of which limit their wider applicability.
Overall, existing literature in this space shows that EKF-based algorithms are susceptible to sensor spoofing attacks.
However, they are either specific to a particular state estimation algorithm, simpler vehicle types, require extensive knowledge about implementation details or require specialized hardware units.

\papername can target state estimation models (\eg EKF) in autonomous systems and generate spoofed input/sensor values to lead the systems astray. 
\papername presents a stealthy attack that is \textit{opaque} \wrt the target function --- \ie there is no knowledge required about the internal details of the state estimator (\ie the target function) --- the only requirement is that the state estimation function be ``learnable'' (explained in \S\ref{sec:dis:learn}) from the observation of its inputs and outputs\footnote{As we see later in \S\ref{sec:threat_model}, this is a very lightweight requirement and doesn't need anything beyond a simple user-level access on the system or communication via middleware.}.
The final result of a \papername-based attack is {\em a deviation of the physical system's trajectory} while the system itself doesn't notice anything untoward (and believes that it is following the original mission parameters).

\papername{}
works via \ca creating a ``surrogate'' model of the target function and \cb training a ``spoofer'' against it.
``Surrogate'' models (often more than one as detailed in \S \ref{sec:req:sm}) try to emulate the target functions (\ie state estimators and anomaly detectors)\footnote{\note we use ``target function'' and ``state estimator'' interchangeably in this paper.} based on observed I/O behavior.
A surrogate model is a deep neural network (DNN) \cite{bishop1995neural} that can capture both, the operational capabilities as well as the vulnerabilities of the target model; hence, the \textit{vulnerabilities that are in the target can be transferred to the surrogate}.
The use of DNNs allows us to develop a ``spoofer''by optimizing against the surrogate similar to generative adversarial networks (GANs) \cite{gan} such a way that
\textit{influence the movement of the UAV}, while \textit{evading detection}.
\note we are not trying to take direct control of the vehicle, rather \textit{influence} it to stray away from the mission parameters, often in a \textit{specific} manner.
The spoofer, once trained, can be used to generate malicious inputs that can \textit{transfer to real drones}.

\papername{} is able to cause \textit{deviations across tens of metres} for different mission scenarios, as shown in both, \textit{real hardware drones in our lab} (built using commodity hardware and open source software) as well as widely-used simulation engines (Gazebo) --- all while evading detection. 
In fact, \papername{} is able to evade the state-of-the-art EKF-based anomaly detector (SAVIOR~\cite{quinonez2020savior}); our approach succeeds in stealthy attack demonstrate up to $2.5m$ of deviation (\S\ref{sec:eval:savior}) in this case.

{Our central thesis is that learnability is an inherent and exploitable property of state estimators designed for predictability; we demonstrate this through \papername{}, a framework that exploits this property via the following contributions:}
\begin{enumerate}
\itemsep0pt
\item a novel, stealthy attack approach aimed at state estimation algorithms for UAVs and \papername{}, a \textit{software-only framework} that implements said approach,
\item demonstration of the attacks on a widely used, real-world flight controller, \viz PX4, for: \ca real hardware drones and \cb a software-in-the-loop (SITL) framework that works with a widely-used simulation engine (Gazebo).
\item successful attacks against the state-of-the-art EKF-centric anomaly detection method (SAVIOR) \footnote{We performed this evaluation based on great feedback from previous review cycle.}
\item insights into when deterministic safety critical systems become learnable and thus vulnerable to ML-driven attacks
\end{enumerate}
We focus on EKF in this paper since it is one of the most common state estimation algorithms in use.
We assume that {\em ``k-Lipschitz''} property in state estimators~\cite{hermann1977nonlinear, huang2008genearalized, castellanos2004limits, allan2021nonlinear} allows for ``learnability'' (\S\ref{sec:dis:learn}).
\papername{}, to the best of our knowledge, is the first attempt at
\ca exploiting such properties of EKF in UAVs, 
\cb doing so while avoiding detection and 
\cc demonstrating the attacks on \textit{real hardware}. 
For our approach to succeed, several steps must come together:
\ci gaining enough understanding of the flight firmware to access (via existing vulnerabilities) the I/O information in order to ``learn'' the estimator function(\S\ref{sec:threat_model}),
\cii collecting comprehensive training data that captures diverse vehicle kinematics~(\S\ref{sec:req:dc}),
\ciii constructing surrogate models that emulate the target function~(\S\ref{sec:req:sm}) from the collected data,
\civ optimizing a spoofer against such a surrogate~(\S\ref{sec:req:sg}),
\cv demonstrating that attacks transfer from the surrogate to the actual system in simulation~(\S\ref{sec:rq:success}),
    as well as physical hardware~(\S\ref{sec:rq:realworld}).
The \papername code has been open-sourced and is available online\footnote{\paperwebsite{}}.
We will first discuss our system model, assumptions, as well as some required background material next.

\section{Background and System Model}
\label{sec:background_and_system}

An autonomous UAV, especially with long ranging missions, often operates beyond the operator's line of sight, relying on the UAV's {\em reported} trajectory for monitoring.
As mentioned in \S \ref{sec:introduction}, our \textit{goal is to avoid detection}, either by the system or the operator.
If our adversary makes changes that are obvious and not stealthy, there is a very good chance that the mission parameters may be adjusted or a new UAV may be launched by the operators. 
We now present an overview of a typical UAV's operational sequence-of-events, system model, state estimation and anomaly detection methods.

\begin{figure}[t]
    \begin{minipage}{\columnwidth}
        \includegraphics[width=\linewidth]{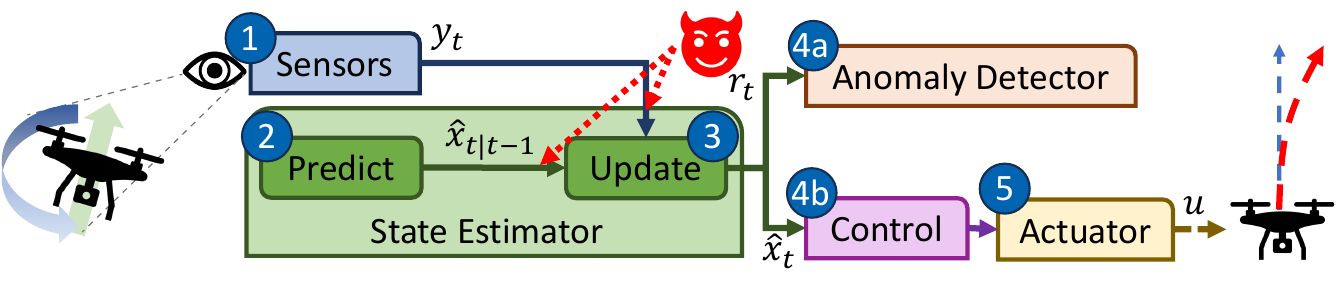}
        \includegraphics[width=\linewidth]{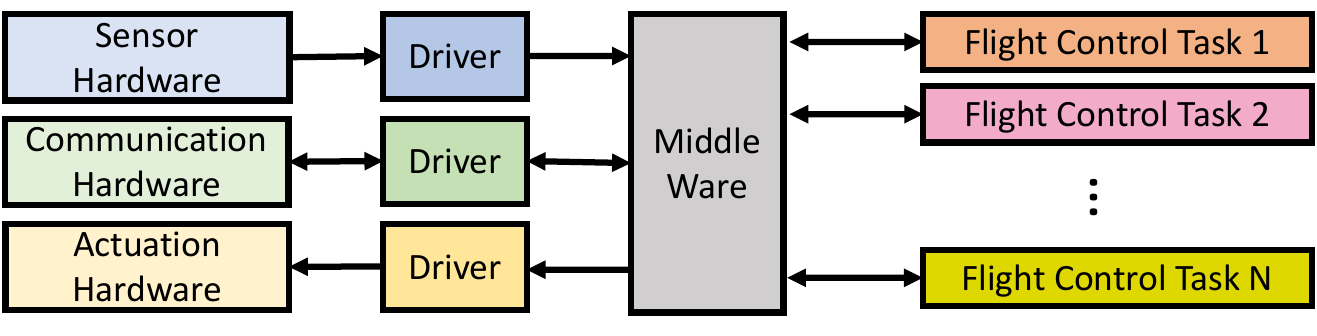}
        
        \caption{\label{fig:sequence_of_events}(Top) Sequence of events during the operation of a UAV.
        (Bottom) Standard software architecture of common flight controllers. %
        }
        
    \end{minipage}
    \vspace{-\baselineskip}
\end{figure}

A state estimator operates as follows (\fig{\ref{fig:sequence_of_events}} top).
\circledSeq{1} Sensors capture the vehicle's kinematics.
\circledSeq{2} The state estimator predicts the current state from the previous actuation.
\circledSeq{3} The sensor values are compared against the prediction.
The differences (\textit{residuals}) are sent to the anomaly detector, the updated state to the control.
{\circledSeqSmaller{4a}} High residuals trigger an alarm and the sensor is rejected.
{\circledSeqSmaller{4b}} Based on the updated state, the control algorithm calculates the actuation commands needed to keep the vehicle within the mission parameters.
\circledSeq{5} The commands are converted to pulse-width modulation (PWM) signals for the motors, resulting in the mission trajectory shown in blue.
The motors affect the kinematics and the cycle continues.

However under the false data injection (FDI) attacks, an adversary can observe the predicted state and manipulate the sensor values (\ie inputs to the update function) before \circledSeq{3}, causing actual trajectory (red) to diverge from the intended path (blue dotted line) while the vehicle perceives nominal operation.
We discuss the threat model in the next section elaborating further about an attacker's capability to achieve such results.

\subsection{Our System Model and Assumptions}
\label{sec:basm:sm}

The system uses middleware
(\fig{\ref{fig:sequence_of_events}} bottom)
to disseminate sensor information across tasks.
Autonomous systems commonly utilize middleware as a software system architecture due to the benefit of being able to add new components to the system without affecting the existing functionality, for instance in robotics (\eg Robot Operating System (ROS)\cite{ros}) and in flight controllers (\eg PX4 \cite{meier2015px4}).
The publish-subscribe architecture allows each components to publish outputs and subscribe for inputs, allowing the tasks to run only when their inputs are available.
Generally, each task only perform a very specific function. Therefore, we assume that a task that executes the target function only runs the target function (\ie no other task executes the target function) {\em without} root access.

We assume a \textit{more restrictive and secure system}\footnote{\label{fn:ad_restrictive}It is more restrictive for the {\em attacker}, therefore, harder for {\em our} objectives.} than traditional flight controllers ---
the latter assumes that all tasks share the same user space environment and have fewer security protocols.
In this paper, we assume that the tasks are isolated in user space, communicating only via a middleware.
Additionally, the integrity of the critical calculations/functions (\eg anomaly detection and control) are \textit{protected} using techniques such as triple redundancy modules~\cite{lyons1962use} or trusted execution environments such as Arm TrustZone~\cite{pinto2019demystifying}, Intel SGX~\cite{intelsgx} or SecureCore~\cite{securecore}, preventing the attacker from changing the logic.
{Note that these protect the integrity of computation but do not inherently protect data in transit between components.}
Therefore, in this paper, we assume that the state estimation's update function is protected; the function logic is unknown to the attacker.

\psection{Sensors}
The system includes Inertial Measurement Unit~(IMU), Global Navigation Satellite System~(GNSS), magnetometer and barometer from autopilot hardware (\eg Pixhawk \cite{priandana2020development}, Navio2 \cite{navio2}).
Although cameras can be used to track movement (\eg optical flow\cite{beauchemin1995computation}), we consider scenarios with low visibility (\eg fog, smoke) or night time operations, thus making the camera useless.
IMU is used to predict the altitude, velocity and position.
GNSS provides the position and velocity~\footnote{As explained in \S\ref{sec:rq:realworld}, due to physical and geographic limitations, we use the ViCon positioning system~\cite{vicon_valkyrie} in our hardware experiments.}.
Magnetometer measures the orientation and barometer measures the relative altitude of the vehicle.
The sensors are used upon reception to avoid outdated values.

\psection{State Estimation}
{
A state estimator, by definition, never knows the true state of the system --- it can only approximate it based on evidence observed from sensors.
This is not a limitation of any particular implementation; it is a property of estimation under uncertainty and, as we demonstrate, it is precisely this property that creates the adversarial space \papername{} exploits.
}
Due to the space constraints, we explain the general process of state estimation  here --- the details of the 
de facto standard for UAVs (\ie Extended Kalman Filter~(EKF) \cite{kalman1960new}) are in \app{\ref{appendix:ekf}}.

State estimation generally has two stages: \textit{predict} and \textit{update}.
The first stage ``predicts'' the future state of the system since sensors may not refresh fast enough or may have accumulated errors.
When the sensor values arrive, the predictions are corrected via \textit{sensor fusion} (\ie the state update step
\footnote{Since the focus this paper is on the state update function, $F$, we will use ``state estimator'' and ``state update'' interchangeably.
}) (\eq{\ref{eq:state_est}}):

\noindent\begin{minipage}{\columnwidth}
\vspace{5mm}
\begin{equation}
    \label{eq:state_est}
    \eqnmarkbox[blue]{update}{\hat{x}_t},
    \eqnmarkbox[red]{residual}{r_t} = 
    F(
    \eqnmarkbox[ForestGreen]{pred}{\hat{x}_{t|t-1}},
    y_t,
    \eqnmarkbox[gray]{args}{\theta{}_t})
\end{equation}
\annotate[yshift=0.5em]{right, above}{residual}{residual}
\annotate[yshift=0.5em]{left, above}{update}{estimated state}
\annotate[yshift=0.5em]{right, above}{pred}{prediction}
\annotate[yshift=-0.25em]{left, below}{args}{implementation specific parameters}
\vspace{2mm}
\end{minipage}

The objective of $F$ is to minimize the \textit{estimation error}, $\epsilon_t =|x_t-\hat{x_t}|$, where $x_t$ is the ground truth (which the estimator \textit{never} knows) at time $t$.
The prediction, $\hat{x}_{t|t-1}$, enables $F$ to find the residual, $r_t$ (differences between observed sensor, $y_t$, and the expected/predicted value),
which indicates anomalous behavior (\ie larger the $|r_t|$ suggests higher anomaly likelihood).
The parameter, $\theta{}$, accounts for various implementations or formulations of the state update function, $F$.
We model $F$ as a \textit{pure function} by lifting its mutable state into the input, enabling the learning of its input-output mapping.
{\em This assumption is critical --- \textit{the very determinism that makes state estimators analyzable/predictable is the same property that makes them ``learnable'' by an adversary}.}

\psection{Learnability}
\label{sec:dis:learn}
For a target function to be \textit{learnable}, it must be differentiable or have bounded continuity:
current ML theory assumes \textit{k-Lipschitz continuity}~\cite{von2004distance, davis2020stochastic, neyshabur2017exploring} that places upper bounds on how fast a function output can change \wrt the input.
Differentiability and the bounded continuity also apply for state estimator analysis to provide guarantees about its behavior \cite{hermann1977nonlinear, huang2008genearalized, castellanos2004limits, allan2021nonlinear}.
Safety critical systems (\eg autonomous vehicles or avionics) require predictability guarantees.
Hence, \textit{our approach exploits the analyzability and inherent determinism of the estimation algorithms}.

Furthermore, our approach is slightly more relaxed: the continuity constraint is \textit{only a subset of the input domain reflective of the vehicle's mission} (\ie locally Lipschitz). %
This is because the state estimator will only observe regions of the input domain corresponding to the deployment mission throughout the attack.

\subsection{Anomaly Detector}
\label{sec:threatmodel:ad}
An anomaly detector (AD) highlights abnormal UAV behaviors (\ie deviation from expectation) using residuals.
The detector must balance between false negatives (\ie be too lax) and false positive (\ie alarm fatigue).
$\chi^2$ \cite{mo2010false} is a statistical AD method often deployed in CPS \cite{jovanov2019relaxing, stealthy_optimal_myopic} and PX4, formulated as \eq{\ref{eq:chisq}}:

\begin{minipage}{0.95\columnwidth}
    \vspace{1.5em}
    \begin{equation}
        \label{eq:chisq}
        r_t^T
        \eqnmarkbox[red]{C}{C^{-1}}
        r_t =
        \eqnmarkbox[blue]{z}{z_t}
        \geq
        \eqnmarkbox[ForestGreen]{eta}{\eta} \rightarrow \frac{z_t}{\eta} \geq 1
    \end{equation}
    \annotate[yshift=0.5em]{right, above}{C}{inverse residual covariance}
    \annotate[yshift=0em]{left, below}{z}{normalized residual}
    \annotate[yshift=0em]{right, below}{eta}{$\chi^2$ threshold}
    \vspace{0.5em}
\end{minipage}
normalized residual, $z_t$ (assumed $\chi^2$ distributed), is compared to user-specified threshold $\eta$ (in standard deviations);
we refer to $\frac{z_t}{\eta}$ as $\chi^2$ score\footnote{\label{fn:ad_chi_bypass}We show \papername{} can easily bypass PX4's $\chi^2$ with default $\eta=5$ in \S\ref{sec:eval}}.
In addition to $\chi^2$, we also evaluate a more stringent threshold-based AD, $\tau$-AD (\eq{\ref{eq:threshold}}), which flags residuals as anomalies when:

\noindent{}\begin{minipage}{0.95\columnwidth}
    \vspace{5mm}
    \begin{equation}
        \label{eq:threshold}
        \eqnmarkbox[red]{residual}{r} \leq
        \eqnmarkbox[blue]{tmin}{
        \tau_{min}} || r \geq
        \eqnmarkbox[blue]{tmax}{
        \tau_{max}}
    \end{equation}
    \annotatetwo[yshift=0.5em]{left, above}{tmin}{tmax}{anomaly threshold}
    \annotate[yshift=0.5em]{left, above}{residual}{residual}
    \vspace{-2mm}
\end{minipage}

The thresholds are determined from the distributions of residuals during nominal missions.
This is more restrictive\footnoteref{fn:ad_restrictive} than CUSUM used by SAVIOR \cite{quinonez2020savior} which uses cumulative past residuals.
We show that \papername{} can bypass $\tau$-AD as well.

\section{Threat Model}
\label{sec:threat_model}
\begin{figure}
    \centering
    \begin{minipage}[b]{0.32\columnwidth}
        \begin{subfigure}{\linewidth}
            \includegraphics[width=\linewidth]{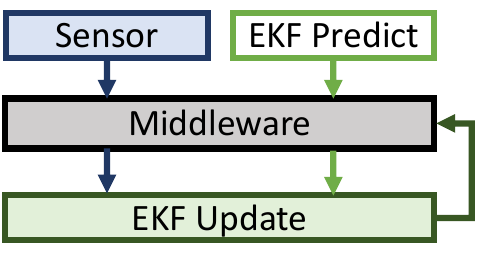}
            \caption{Nominal data flow}
            \label{fig:ekf_task}
        \end{subfigure}
    \end{minipage}
    \begin{minipage}[b]{0.32\columnwidth}
        \begin{subfigure}{\linewidth}
            \includegraphics[width=\linewidth]{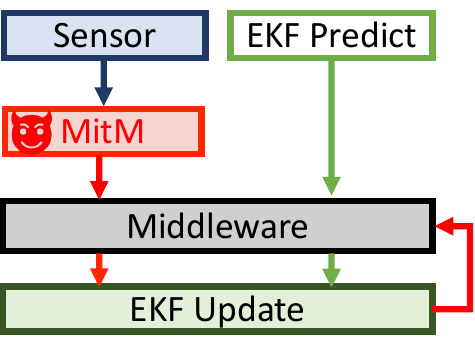}
            \caption{Topic Hijacking}
            \label{fig:ekf_task_mitm}
        \end{subfigure}
    \end{minipage}
    \begin{minipage}[b]{0.32\columnwidth}
        \begin{subfigure}{\linewidth}
            \includegraphics[width=\linewidth]{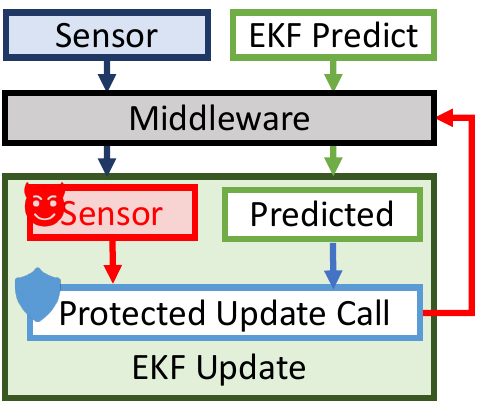}
            \caption{Input corruption}
            \label{fig:ekf_task_bad_input}
        \end{subfigure}
    \end{minipage}
    \caption{\label{fig:atk_entry}Examples of adversary entry vectors.
    {
    The entry vector depends on how the firmware is designed.}
    (\subref{fig:ekf_task}) shows the nominal flow.
    (\subref{fig:ekf_task_mitm}) is an example of a malicious module in the system hijacking a sensor topic
    --- this vector can be used when all of the inputs and outputs needed for an update are passed via middleware (\eg \cite{teixeira2020security, jeong2017study, dieber2020penetration}), while the spoofer inference runs as a separate module.
    (\subref{fig:ekf_task_bad_input}) is an example of the attacker exploiting vulnerabilities in tasks to corrupt the inputs (a common entry mechanism in the security world, \eg \cite{cisabufferoverflow,cisamemorysafety,cisaroadmap}) --- 
    {this method is used when not all inputs are available and spoofer inference executes within a userspace module.}
    }
\end{figure}

The attacker's (\ie our system's) main objective is to alter vehicle movements while the operation \textit{seems} nominal without the anomaly detector raising any alarms.
Hence, it is in \textit{the attacker's best interest \textbf{not} to crash the vehicle or be detected too soon}.

Our threat model is similar to ARES~\cite{ding2023get}
\footnote{They use reinforcement learning with knowledge of implementation whereas we
don't have such knowledge.
Also we are more focused on stealth.}
where we assume: attackers collect the target I/O values offline and observe/inject inputs online.
We \textit{broaden this assumption} by treating the target function/module as {\em opaque} --- \ie assuming visibility of only its inputs and outputs rather than its internal logic; this actually handicaps our system more than ARES and other similar work.
We assume realistic attack vectors (explained later in this section).
Literature shows that multi-vendor avionics development practices (even for unmanned systems) \cite{horng2006comparative, rashid2009vendor, li2018supply}
can enable such access via
\ca inconsistent security practices \cite{aerospace_supply_chain,mohan2014real,pellizzoni2015generalized}, 
\cb latent bugs in the system and
\cc potential combination of commodity/open-source components and proprietary software (\eg the flight controller),
allowing access to a copy of the component or offline system replication.
\textit{We also assume that the target function/module does not need root access}.

As a result, our threat model is more realistic than prior work~\cite{kim2019rvfuzzer, choi2018detecting, foruhandeh2019simple}:
\ci{} we assume attackers have additional avenues of exploiting middleware security vulnerabilities, as shown in \fig{\ref{fig:atk_entry}}, (\eg message man-in-the-middle (MitM) attack~\cite{jeong2017study, dieber2020penetration, teixeira2020security}) on top of firmware exploitation (\eg memory corruption~\cite{maggi2017rogue, domin2016security}, buffer overflow~\cite{mavr}, \etc),
\cii{} our model anticipates a security architecture (\S \ref{sec:basm:sm}) where \textit{critical flight logic function is sandboxed or isolated} (\fig{\ref{fig:ekf_task_bad_input}}), but I/O remains accessible --- a common pattern in modular avionics~\cite{horng2006comparative, rashid2009vendor, li2018supply},
\ciii{} there are existing methods~\cite{schedule_leak, liu2019leaking} that use the timing behavior to find the target function and
\civ{} the certification process~\cite{jacklin2012certification, hochstrasser2019application} involves integration testing, often using simulations, enabling offline estimation/capture of the I/O behavior, expectations and system configuration.
{An attacker could chain these vectors by,
\ca first replicating the system offline to collect I/O,
\cb training the surrogate and spoofer using that collected information and then 
\cc deploying the trained spoofer via one of the aforementioned vectors at runtime.}

{\note \papername{}'s formulation doesn't specify a specific delivery mechanism, requiring only observability of the target function's I/O and timely injection of inputs.
This paper \textit{focuses on software-based attack vectors}. 
Hardware-based/physical sensor spoofing attacks are \textbf{beyond the scope of this work} since they require \textbf{physical delivery mechanisms} (\ie GNSS signal spoofing~\cite{kerns2014unmanned, tractorbeam, stealthy_gps_navigation, fusionripper, kim2024systematic}, acoustic IMU spoofing~\cite{attack_vec_imu_sound, walnut, poltergeist, kim2024systematic}, \etc) or line-of-sight access --- all of which require additional hardware, physical proximity, precise timing and attack/target configurations --- a very different set of mechanisms.

}

\section{Problem Statement}
\label{sec:problem_statement}

Our objective is stealthy deviation \ie increasing estimation error without triggering anomaly detection until the \textit{mission is {irreversibly} compromised}
--- \textit{hence, the drone can neither distinguish modified sensor inputs from correct ones nor detect position drift, leaving little to no opportunity for recovery by the time the deviation becomes apparent}.
Prior work \cite{kim2019rvfuzzer, choi2018detecting, foruhandeh2019simple} requires implementation knowledge of the vehicle's estimation algorithm or reverse engineering the logic to achieve this through sensor manipulation alone.
Per \eq{\ref{eq:state_est}}, the UAV uses state update, $F$, to calculate residuals and new state estimations.
In our case $F$ can be queried offline\footnote{Querying $F$ is optional as we show real-world results without query access.},
but the attacker faces the following constraints: unknown algorithm type, unknown deployment mission, offline access to inputs ($y_t$, $\hat{x}_{t|t-1}$, $\theta$) and outputs ($r_t$, $\hat{x}_t$), online input visibility, but only online manipulation capability for sensor subset $y_t$. 
Therefore, our problem statement is:
\begin{itemize}
    \item How to mimic $F$ only using inputs and outputs?
    \item How to apply this knowledge to create adversarial inputs?
\end{itemize}

\noindent\papername{} addresses these problems by:
\begin{itemize}
    \item Building a {\em surrogate model} $\hat{F}$ that locally approximates $F$ from input-output data
    \item Optimizing a {\em spoofer} via gradient-descent on the surrogate to minimize the residuals (anomaly indicator variable) while maximizing estimation error.
\end{itemize}

Building \papername{} is therefore an {\em offline process} using representative `mock' missions {\em instead of} 'live missions'
--- valid assumptions from prior work \cite{ding2023get} and existing exploits (\S\ref{sec:threat_model}). %
\noindent We describe \papername{}'s design in the next section.

\section{\paperName}
\label{sec:requiem}
Unlike analytical control-theoretic attacks which require manual derivation of the specific vehicle's dynamics and estimator matrices (which may be unknown or proprietary), \papername{} automatically learns the necessary spoofing policy solely from I/O observations. This makes it vehicle-agnostic.

\papername{}'s objective --- stealthily increasing the estimation error to cause physical deviations --- hinges on \ca the surrogate model implicitly learning the state update function $F$'s design faults and \cb the spoofer exploiting them.
This raises three design questions:
\begin{enumerate}[label=\dqu{\arabic*},left=2pt]
    \item \label{dq:data} How to effectively get data to construct a surrogate?
    \item \label{dq:overfit} How to prevent overfitting and underfitting?
    \item \label{dq:spoofer} How to optimize the spoofer to be effective?
\end{enumerate}

\begin{figure*}
    \centering
    \begin{minipage}{\textwidth}
        \includegraphics[width=\textwidth]{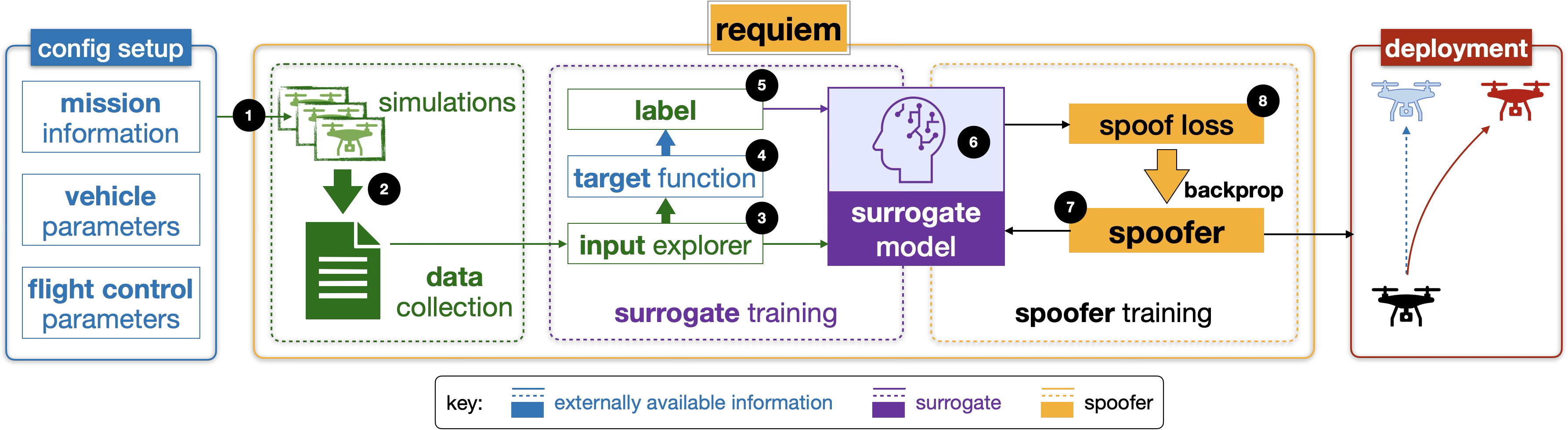}
        \caption{\papername Pipeline.
        \circledReq{1} Simulations are configured to best reflect a deployment scenario. 
        \circledReq{2} data is collected over multiple simulation runs.
        \circledReq{3} generate more data similar to those collected.
        \circledReq{4} generated data is sent to the target function to \circledReq{5} get the output (\ie label). 
        Input-output pairs are used to \circledReq{6} train the \textit{surrogate} model.
        \circledReq{7} a spoofer model is then trained to optimize against the surrogate where the loss function \circledReq{8} provides feedback to the spoofer, shaping its attack behavior.
        }
        \label{fig:main_pipeline}
    \end{minipage}
\end{figure*}

Our three-stage pipeline (\fig{\ref{fig:main_pipeline}}) addresses these questions:
data collection (\S\ref{sec:req:dc}), surrogate training (\S\ref{sec:req:sm}), spoofer training (\S\ref{sec:req:sg}).
\S\ref{sec:req:imp} describes our implementation for PX4.

\subsection{Data Collection}
\label{sec:req:dc}
We leverage a flight controllers' simulation interfaces\footnote{Even commercial, proprietary flight controllers provide hardware-in-the-loop simulation functionality \cite{dji_osdk}\cite{perezsimulation}\cite{parrot_sphinx}.}
used for certification~\cite{jacklin2012certification}\cite{hochstrasser2019application}
to collect surrogate training data.
Although physical UAV operation could provide this data, simulation is more convenient.
The simulation collects a set of inputs $S$ (\eq{\ref{eq:snapshots}}).
If $F$ can be queried, outputs are obtained directly; otherwise both inputs and outputs must be collected.

\begin{equation}
    \label{eq:snapshots}
    S:=\{s_0, ..., s_{T}\}, s_t:=\{\hat{x}_{t|t-1}, y_t, \theta_t\} \forall t\in T
\end{equation}

\noindent\textbf{\ref{dq:data} \label{sec:dq:data}: How to effectively collect data to construct a surrogate?}
Without knowing the deployment mission, effective data collection requires capturing diverse vehicle kinematics to cover various scenarios. 
We use a random walk mission (\app{\ref{appendix:missions}}) where each setpoint is a fixed distance from the previous setpoint in a random direction and random heading.
This captures diverse lateral linear movements and heading changes, but 
excludes circular movements ---
a constraint we exploit to test \papername{}'s robustness against unexpected kinematics (\S \ref{sec:eval}).

\subsection{Surrogate Model}
\label{sec:req:sm}

Surrogate $\hat{F}$ (a DNN) emulates $F$ to transfer attacks from $\hat{F}$ to $F$ via gradient descent.
The parameters of $\hat{F}$ are adjusted to {\em minimize} the output differences between $\hat{F}$ and $F$ (\eq{\ref{eq:surrogate_objective}}) using loss
$\mathcal{L}$ (\eg mean squared error, Huber Loss~\cite{huber1992robust}).

\noindent{}\begin{minipage}{\columnwidth}
\begin{equation}
    \label{eq:surrogate_objective}
      \min_{\hat{F}}
      E\big[\mathcal{L}({
        \eqnmarkbox[DarkOrchid4]{surrogate}{\hat{F}(s)},
        \eqnmarkbox[Blue]{target}{F(s)})
        }\big], \forall s \in S
\end{equation}
\annotate[yshift=-0.25em]{left, below}{surrogate}{surrogate output}
\annotate[yshift=-0.25em]{right, below}{target}{target output}
\vspace{2mm}
\end{minipage}

\noindent\textbf{\ref{dq:overfit}: How to prevent overfitting and underfitting?}
Deep models risk overfitting with insufficient data or underfitting when $F$ is too complex.
While increasing $\hat{F}$'s complexity could address underfitting, it demands more data and computation.
Therefore, we use a simple \textit{fully connected feed-forward neural network} and address both issues via \ca \textit{data augmentation} to increase diversity of training data and \cb \textit{slicing} to reduce complexity by emulating specific aspects of $F$.

\psection{Data Augmentation}
Data augmentation generates new data without additional collection to improve generalization (\ie reduce overfitting).
We use Monte Carlo Arithmetic \cite{stott1997monte, kiar2021data}
to generate new data and reduce overfitting by adding noise from random distribution to the inputs.
The objective function for $\hat{F}$ is modified to \eq{\ref{eq:surrogate_obj}} where the bound of the uniform distribution ($w_l,w_u$) are set to respect domain constraints (\eg heading angle $> 2\pi$ rad) and 
$E[\cdot ]$ denotes expected value (\ie mean).
This expands the domain where $\hat{F}$ can emulate $F$.

\noindent\begin{minipage}{\columnwidth}
    \centering
    \vspace{4mm}
    \begin{equation}
        \label{eq:surrogate_obj}
        \begin{aligned}
            \forall \eqnmarkbox[Green4]{snap}{s}\hspace{-1mm}\in\hspace{-1mm}{}S
            ,
            \eqnmarkbox[Red]{noise}{\eta}\hspace{-1mm}\sim{}\hspace{-1mm}
            \eqnmarkbox[DarkSlateGray4]{uniform}{U}(\omega_{l},\omega_{u})\\
            \min_{\hat{F}}\bigg[ E\big[
            \mathcal{L}(
            \hat{F}(s+\eta),
            F(s+\eta))\big]\bigg]
        \end{aligned}
    \end{equation}
    \annotate[xshift=0.5em, yshift=0.5em]{left, above}{noise}{noise}
    \annotate[yshift=0.5em]{right, above}{uniform}{uniform dist.}
    \annotate[yshift=0.5em]{left, above}{snap}{input}
\end{minipage}

\psection{Slicing} To avoid underfitting, we reduce $F$'s complexity by limiting the number of variables $\hat{F}$ must emulate --- \ie $F$ emulates a {\em slice} of $F$ or multiple $\hat{F}$s to emulate multiple slices.
For example, an attacker manipulating only GNSS doesn't need magnetometer residuals since they measure different aspects.
Therefore, we can create a surrogate that only emulates the GNSS residual calculation (\ie {\em slicing} $F$ into GNSS residuals) as shown in \eq{\ref{eq:slice}}.
Slicing prevents $\hat{F}$ from optimizing for output variables that are not of interest.

\noindent{}\begin{equation}
    \label{eq:slice}
    F: s_ \rightarrow
    (\hat{x}, r^{GNSS}, r^{mag},
        \eqnmarkbox[DarkSlateGray4]{other}{\psi}) ,
        \eqnmarkbox[Blue]{slice}{F_{[GNSS]}} : s \rightarrow (r^{GNSS})
\end{equation}
\annotate[yshift=-0.25em]{left, below}{other}{other output variables}
\annotate[yshift=-0.25em]{right, below}{slice}{sliced into GNSS residual}

\subsection{Spoofer}
\label{sec:req:sg}
The spoofer model, $G$, is a DNN that is trained using gradient-based optimization against the surrogate $\hat{F}$ such that it generates a spoofed sensor value so that the resulting residual remains low (\ie stealthy) and the estimation error increases (\ie deviation) when transferred to the target $F$.
The gradient-based optimization is possible since the surrogate is a DNN.
To keep $G$ computationally light weight for faster inference during deployment but expressive enough to optimize against $\hat{F}$, we use a fully connected feed-forward neural network for $G$.

\noindent\textbf{\ref{dq:spoofer} How to optimize the spoofer to be effective?}
{
    If injected sensor values consistently match the estimator's expectations (\ie low residuals), the estimator never corrects its growing estimation error --- and since anomaly detectors also rely on residual magnitude, low residuals simultaneously ensure stealth.
    Therefore, we propose two strategies exploiting this: No Correction (NC) and Direction Bias (DB), explained below.
}

\psection{No Correction}
In this attack, a spoofer $G$ prevents the estimator from realizing the existence of errors by {\em injecting sensor values that match the estimator's expected sensor value}.
Therefore, $G$ is optimized to {\em minimize the residual} as shown in \eq{\ref{eq:anomaly_loss}}:

\noindent\begin{minipage}[t][1.2cm][t]{\columnwidth}
    \centering
    \vspace{2mm}
    \noindent\begin{minipage}[t][1cm][t]{\linewidth}
        \begin{equation}
            \label{eq:anomaly_loss}
            \min_G \big[ \mathcal{L}(L_a) \big] \hspace{1mm}\text{where}\hspace{1mm}
            L_a=
            \eqnmarkbox[ForestGreen]{R}{R}(
                \eqnmarkbox[Purple]{F}{\hat{F}(s+G(s)))}
        \end{equation}
        \annotate[yshift=0.5em]{left, above}{R}{residual extractor}
        \annotate[yshift=0.5em]{right, above}{F}{surrogate output}
    \end{minipage}
\end{minipage}
\noindent{}$R$ is a function that only returns the residuals.
It can cause the vehicle to behave in a dead reckoning mode.

\psection{Direction Bias}
This attack trades off some residual (\ie stealthiness) to {\em bias the direction of the estimation error} by exploiting the attacker's knowledge about the anomaly threshold to keep residuals within the threshold.
It can cause the vehicle to {\em deviate towards a direction of the attacker's choosing}.
$G$ is optimized to maximize deviation per residual without exceeding the threshold.
Therefore, the optimization is a combination of three types of losses: residual loss (\eq{\ref{eq:anomaly_loss}}, deviation loss (\eq{\ref{eq:deviation_bias_loss}}) and budget loss (\eq{\ref{eq:budget_loss}}).

\textit{Deviation loss} measures how much the estimation changed due to the spoofer as shown in \eq{\ref{eq:deviation_bias_loss}}:

\noindent{}\begin{minipage}[t][1.4cm][t]{\linewidth}
    \vspace{2.5mm}
    \begin{equation}
        \label{eq:deviation_bias_loss}
        L_d=
        ReLU(
        \eqnmarkbox[Blue]{predicted}{P(F(s))} -
        \eqnmarkbox[Orange]{spoofed}{P(\hat{F}(s + G(s)))}
        )\\
    \end{equation}
    \annotate[yshift=-0.25em]{left, below}{predicted}{estimated position}
    \annotate[xshift=0.2em, yshift=0.5em]{left, above}{spoofed}{estimated position resulting from spoof}
    \vspace{2mm}
\end{minipage}

\noindent{}$P$ extracts the position values of the estimation.
$ReLU$\cite{nair2010rectified} is used to ensure that the loss increases only if the spoofed values cause the state to update in favor of a desired direction.
Therefore, the direction of the bias can be adjusted by swapping the two terms inside the ReLU.

\textit{Budget loss} (\eq{\ref{eq:budget_loss}}) is used to ensure the residual does not exceed the anomaly threshold by measuring the ``slack'' between the current residual and the anomaly threshold, $T$.
During training, $T$ should be set below the target anomaly detector's threshold to account for the noise.

\noindent{}\begin{minipage}{\linewidth}
    \begin{equation}
    \label{eq:budget_loss}
        L_b=
    ReLU(T-|L_a|)
    \end{equation}
\end{minipage}

The maximization of $L_d\cdot{}L_b$ results in maximizing deviation {\em per residual}.
Since we still need to incorporate $L_a$ to remain stealthy; we end up with the following objective function shown in \eq{\ref{eq:db_opt}}:

\noindent{}\begin{minipage}{\linewidth}
    \begin{equation}
        \label{eq:db_opt}
        \min_G \big[ (L_a)^2-L_b \cdot{}L_d \big]
    \end{equation}
\end{minipage}

\subsection{Implementation for PX4}
\label{sec:req:imp}
In PX4, we targeted the \texttt{ekf2} module that use the EKF library class, specifically \texttt{controlFusionModes()} that performs the sensor fusion/state update (\ie $F$).
We assume that the attacker can manipulate the {\em north position} and {\em velocity} as reported by GNSS.
Since PX4's state estimation is not a pure function (rather a class method), it can be treated like a pure function by treating the EKF object member variables as an input.
Note that \papername{} is not aware of the internal logic of \texttt{controlFusionModes()}.
Therefore, the collection procedure is implemented by recording all variables in the EKF object (\ie $s_t$ in \eq{\ref{eq:snapshots}}) right before $F$.

There were $2020$ variables and we can use it to query $F$ offline.
However, the majority of variables did not change during the collection as many were static configurations.
Furthermore, $F$ is sliced (\S\ref{sec:req:sg}) into \ci GNSS residuals, \cii position and \ciii velocity estimation since we are attacking GNSS value: the position and velocity estimation are needed to construct the DB attack.
We use the heuristic of eliminating inputs that do not affect outputs by checking for changes in the output after adding large values to the variable question.
In other words, if $F(s) = F(s+\eta)$ where $\eta$ is non-zero for one of the indices, then that index has no effect on the output (\ie the index can be discarded).
Therefore, we managed to narrow down the input size for training \papername{} to variables presented in \tab{\ref{tab:collected_data_type}}.%
\begin{table}[]
        \footnotesize
        \caption{\label{tab:collected_data_type}Types of data collected for PX4 simulation evaluation. The spoof target is highlighted in yellow (\ie GNSS)}
        \vspace{-2mm}
        \begin{tabular}{|p{0.5in}|>{\raggedright\arraybackslash}p{0.75in}|p{1.75in}|}
        \hline
        \cellcolor[HTML]{bdbdbd}\textbf{Type}          & \cellcolor[HTML]{bdbdbd}\textbf{Name}          & \cellcolor[HTML]{bdbdbd}\textbf{Description} \\ \hline
        \multirow{7}{*}{State} & Attitude ($\phi$)             & Represented in quaternion frame\\ \cline{2-3} 
                               & Position ($p)$ and Velocity ($v$)  & Represented in (m) and (m/s) respectively in North, East and Down coordinate frame.\\ \cline{2-3} 
                               & $\Delta$ Angle Bias ($\phi_b$) & Bias in IMU angle measurement in radians\\ \cline{2-3} 
                               & $\Delta$ Velocity Bias $v_b$ & Bias in IMU velocity measurement in (m/s) \\ \cline{2-3} 
                               & Earth magnetic field ($\beta_e$)   &  Represented in  gauss \\ \cline{2-3} 
                               & Body magnetic field ($\beta_b$) & Magnetic field of the vehicle in gauss       \\ \cline{2-3} 
                               & Wind velocity ($v_w$)& Used to calculate drag force typically used in fixed wing vehicles.\\ \hline
        \multirow{2}{*}{Sensors}
            & \cellcolor[HTML]{FFFAA0} GNSS  & Measures $p$, $v$, horizontal and vertical position noise variances, and velocity noise variance\\ \cline{2-3} 
            & IMU  & Measures angular velocity in (Rad/s) \\\hline
        Imp. Specific Args. ($\theta$) & Body-to-earth frame & Rotation matrix that transforms from body frame to earth frame.\\\hline
        \end{tabular}
\end{table}

We ran 20 separate missions for data collection
that were split into training set and validation set as follows:

\begin{tabular}{|c|c|c|c|}\hline
      & Missions & \# of Samples & After Augmentation\\\hline
     Train &  12 & 2275 & 79625\\\hline
     Val. & 8 & 1587 & 55545\\\hline
\end{tabular}

\noindent{}We trained three surrogate models (one for each slice) and two spoofer models (NC and DB) using PyTorch with Adam optimizer \cite{kingma2014adam}.
Due to the space constraints, the details of the surrogate and spoofer model training implementations are elaborated in \app{\ref{appendix:implementation}}.
We describe the test procedure in the next section.

\section{Evaluation and Results}
\label{sec:eval}

A successful stealthy attack must show that the vehicle is operating in a {\em seemingly} nominal state while causing the actual  trajectory to deviate.
We perform all of the evaluation missions but will discuss only some of the example here due to the space limitations\footnote{All of the missions that were performed are corresponding results and data have been uploaded to our website \paperwebsite{}}. 
We outline our evaluation methods and results in this section to show that \papername{} successfully carries out stealthy attacks.
Evaluating the effectiveness of \papername{} raises the following research questions (RQ):
\begin{enumerate}[label=\rqu{\arabic*},left=2pt]
    \item \label{rq:success} (\S \ref{sec:rq:success}) To what extent does \papername{} succeed?
    \item \label{rq:realworld}(\S\ref{sec:rq:realworld})
    How well does the attack work on real-world?
    \item \label{rq:weather}(\S\ref{sec:rq:weather})
    How do environmental factors affect the attack?
    \vspace{1mm}
    \hrule
    \vspace{1mm}
    \item \label{rq:stealthy}(\S \ref{sec:rq:stealthy}) What variables affect an attack's ``stealthiness''?
    \item \label{rq:meaningful}(\S \ref{sec:rq:meaningful}) What factors make attacks to be "meaningful"?
    \item \label{rq:learnable}(\S\ref{sec:rq:learn})
    How well did surrogate models \textit{"learn"} a target function? 
\end{enumerate}
The first 3 RQs are quantitative while the last three are more qualitative.
We first define the evaluation metrics, experiment parameters and processes used to answer each of these RQs.

\subsection{Metrics}
\label{sec:eval:subsec:metrics}
\noindent
We evaluate four trajectories: Planned (ideal), Nominal (no attack), Attack (actual path under spoofing), and System POV (estimator output).
Larger the difference between {\em attack} and {\em system POV} trajectories, larger the deviation.
We must also consider ``stealthiness'' when evaluating these attacks where the attack's success depends on achieving \ca {\em seemingly} nominal operations of the vehicle (\ie \ref{rq:stealthy}) and \cb significant deviation (\ie \ref{rq:meaningful}) {\em simultaneously}.
Therefore, we provide metrics for the attack \wrt the anomaly detector response and deviations.

As mentioned in \S \ref{sec:threatmodel:ad}, two types of anomaly detectors are considered: \ci {\em onboard AD} (\ie $\chi^2$ \S \ref{sec:basm:ekf}) and the more stringent \cii {\em threshold AD} (\ie $\tau$, \S \ref{sec:threatmodel:ad}).
The threshold of $\tau$-AD was set to upper and lower 2.5 percentile (\ie $\tau_{max}, \tau_{min}$)
\footnote{
Conservative bound since the bounds need to be more lax to achieve zero false positives under a nominal mission -- \ie we used a tighter (more restrictive) bound for \textit{our} system since the default values were too easy to spoof
}
of the nominal residual distribution for each mission. 
A stealthy attack mainly cares about whether the anomaly detector raises an alarm during a mission and the deviations achieved without triggering.
\tab{\ref{tab:metrics}} describes the corresponding metrics, \viz:

M1/M3 indicate detection (overtness), while M2/M4 measure maximum deviation before detection.
An attack is considered \textit{partially successful} if the attack becomes overt at a much later stage.
However if an attack becomes \textit{overt} within a second of the attack, it is considered a failure \wrt the chosen anomaly detector.
In fact, {\em stealthy attacks should not trigger any of the anomaly detectors}.
In a partially successful attack, the ``maximum stealthy deviation'' is the largest deviation until the attack becomes \textit{overt}.
An effective attack should have as much stealthy deviation as possible within a given mission.

The adversary may care more about changing the shape (or path) of the trajectory while retaining the appearance that the UAV is following the planned path in missions (\eg where UAVs are used to survey an area).
Therefore, we must quantify the {\em trajectory shape difference} between the planned path and the system POV trajectory, respectively both compared to the `nominal' paths. 
To do that, we use total Bregman divergence (\TBD{}) \cite{liu2010total} that compares the Gaussian mixture model (GMM) \cite{mclachlan2019finite} of one shape to another.
Due to inherent randomness of GMM, we generate the \TBD{} value 10 times and graph the distribution as box plots.
{\em A good stealthy attack should have a high \TBD{} value for the attack trajectory and a low \TBD{} value for the system POV.}

The metrics will show that \papername{} {\em induces significant deviations while remaining stealthy across various mission scenarios}.
The justification of the metrics \wrt capturing the stealthiness and the significance of the attack are answered by \ref{rq:stealthy} (\S \ref{sec:rq:stealthy}) and \ref{rq:meaningful} (\S \ref{sec:rq:meaningful}) respectively.
Learnability of EKF is answered by \ref{rq:learnable} (\S \ref{sec:rq:learn}).
The next subsection specifies the evaluation parameters and the process to show that \papername{} is necessary for stealthy attacks.

\subsection{Evaluation Parameters}
\label{sec:eval:subsec:exp_param}

We design the experiments to show the effectiveness of \papername{} as a stealthy attack under various mission scenarios.
This subsection establishes the baseline attacks and missions that we will use to evaluate \papername{}.
To demonstrate the feasibility of \papername{}, we test our approach in real drones and in widely used simulation with an open-source flight controller PX4~\cite{meier2015px4} with gazebo~\cite{gazebo} physics simulation~\cite{amir_stealthy_perception, ding2023get, garcia2023ros, rivera2019unmanned}.

\psection{Baseline Attacks}
    To show that \papername{} is necessary, we establish baseline attacks as attack strategies that are agnostic to the input of target function.
    We consider three classes of attacks, specified in \tab{\ref{tab:naive_attacks}}:
    \ci \textit{constant offset}~\cite{van2018veremi, kamel2019catch, kamel2020simulation, kamel2020veremi, hawlader2021intelligent,kamel2020f2md}, 
    \cii \textit{random offset}~\cite{van2018veremi, kamel2019catch, kamel2020simulation, kamel2020veremi, hawlader2021intelligent,kamel2020f2md} and 
    \ciii \textit{boiling frog}~\cite{Erba2023, panda2025real, alhoraibi2024detection} where values for position and velocity are \textit{slowly ramped up} in order to avoid rapid changes and hence, quick detection.
    These baselines are adapted from standard attack models in vehicular security domains~\cite{hasan2020securing}, where offsets of $250m$\cite{van2018veremi}, $70m$~\cite{kamel2020f2md} and $7m/s$~\cite{kamel2020f2md} are typical due to the larger operational scale of ground vehicles.
    We scaled the magnitude down to 1m (position) and 1m/s (velocity) to reflect the tighter operational tolerances of UAVs~\cite{czyza2023assessment}, where deviations of even a few meters can compromise a mission~\cite{skyler2026lightshow, marek2025collision}.
    We use boiling-frog attack, that gradually injects total value of 1 over a minute, to show that even minuscule injections can cause the attack to become \textit{overt}.

\psection{Missions}
    The attack performance is affected by the kinematics of a mission since different missions imply different input distributions of the state estimator.
    To test the attack's robustness against various vehicular kinematics, we specify missions with different trajectories as described in \tab{\ref{tab:simulation_missions}}: circle, linear and hold.

    \begin{table}[t]
        \footnotesize
        \caption{\label{tab:simulation_missions}Mission scenarios with an example mission trajectory and average estimation error over 10 trials, $E[\epsilon]$.
        The exact parameters of the mission are defined in the \app{\ref{appendix:missions}}}
        \footnotesize
\centering
    {
    \def\descWidth{1.7in}
    \def\nameWidth{0.3in}
    \def\trajWidth{0.5in}
    \def\resWidth{1in}
    \def\chiWidth{0.65in}
    \def\errWidth{0.35in}
    \def\tauWidth{1in}
    \begin{NiceTabular}{@{}|p{\nameWidth}|p{\descWidth}|@{}p{\trajWidth}@{}|@{}p{\errWidth}@{}|@{}}
        \hline
        \cellcolor[HTML]{bdbdbd}\textbf{Name} & 
            \begin{minipage}{\descWidth}
                \centering
                \cellcolor[HTML]{bdbdbd}\textbf{Description}
            \end{minipage}& 
            \begin{minipage}{\trajWidth}
                \centering
                \cellcolor[HTML]{bdbdbd}\textbf{Trajectory}
            \end{minipage}
            &
            \begin{minipage}{\errWidth}
                \centering
                \cellcolor[HTML]{bdbdbd} \bm{$E[\epsilon]$}
            \end{minipage}
            \\\hline\hline
        \begin{minipage}{\nameWidth}
            \centering
            Circle
            \circleMission
        \end{minipage}&
        \begin{minipage}{\descWidth}
            Vehicle moves forward with fixed yaw-rate for a specified time. The resulting trajectory forms multiple circles in north and east plane. 
        \end{minipage}
        &
             \begin{minipage}{\trajWidth}
                    \includegraphics[width=\textwidth]{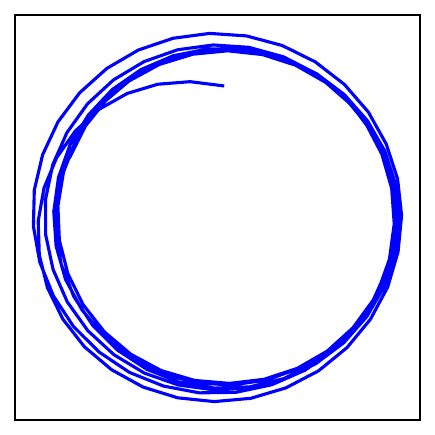}
            \end{minipage}
        &
            0.06147
            \\\hline
        \begin{minipage}{\nameWidth}
            \centering
            Linear
            \linearMission
        \end{minipage}
        &
        \begin{minipage}{\descWidth}
        Setpoints are set along the lateral plane. The resulting trajectory forms a triangle
        \end{minipage}
        &
             \begin{minipage}{\trajWidth}
                    \includegraphics[width=\textwidth]{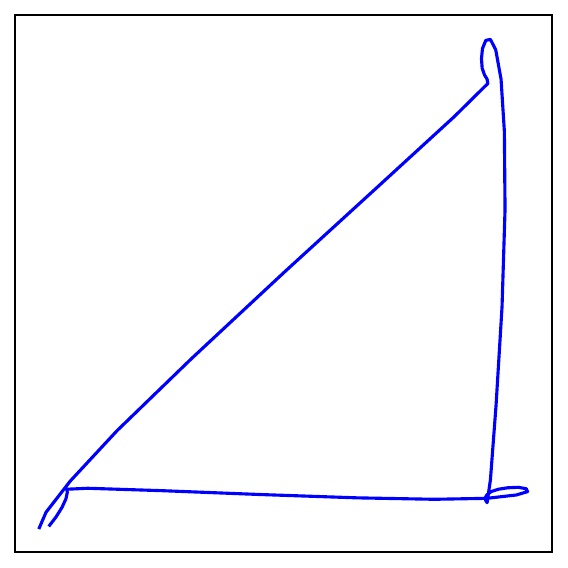}
                \end{minipage}
        &
            0.0418
            \\\hline
        \begin{minipage}{\nameWidth}
            \centering
            Hold
            \holdMission
        \end{minipage}&
            \begin{minipage}{\descWidth}
                 Maintain a specified altitude above the starting point for a specified time.
            \end{minipage}
           &
                \begin{minipage}{\trajWidth}
                    \includegraphics[width=\textwidth]{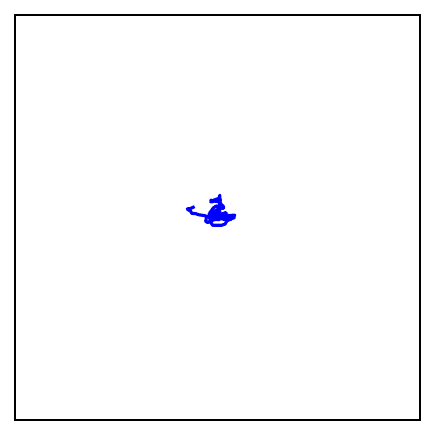}
                \end{minipage}
            &
                0.0159
            \\\hline
        
    \end{NiceTabular}
    }

    \end{table}

    \begin{table}[t]
        \footnotesize
        \centering
        \caption{\label{tab:attack_types} \papername{} stealthy attacks}
        \vspace{-1mm}
            \newcolumntype{P}[1]{>{\raggedright\arraybackslash}p{#1}}
    \begin{tabular}{@{}|P{0.90in}@{}|p{2.3in}|@{}}\hline
        \cellcolor[HTML]{FFD700}\textbf{Name} & \cellcolor[HTML]{FFD700}\textbf{Description}\\\hline\hline
            No Correction (NC) &
            Induces dead reckoning by neutralizing sensor utility
            \\\hline
            Direction Bias (DB) &  The vehicle deviates towards a particular direction\\\hline
    \end{tabular}

        \vspace{2mm}
        \caption{\label{tab:naive_attacks}Baseline attacks. Note: Boiling Frog strategy is the most insidious in this group due to slowly biasing the vehicle.}
        \vspace{-1mm}
        \newcolumntype{P}[1]{>{\raggedright\arraybackslash}p{#1}}
\begin{tabular}{|P{0.35in}|P{0.25in}|p{2.35in}|}\hline
        \cellcolor[HTML]{bdbdbd}\textbf{Type} & \cellcolor[HTML]{bdbdbd}\textbf{Name} & \cellcolor[HTML]{bdbdbd}\textbf{Description}\\\hline\hline
        \multirow{2}{*}{Constant}
                & PCO & Inject a constant value to the GNSS north position\\\cline{2-3}
                & VCO & Add a constant value to the GNSS north velocity\\\hline
        \multirow{2}{*}{Random}
                & PRO & Random value is injected to the GNSS north position\\\cline{2-3}
                & VRO & Add a random value to the GNSS north velocity\\\hline
        \multirow{2}{0.3in}{\centering \textit{Boiling} \newline \textit{Frog}}
                & \cellcolor[HTML]{dddddd}PBF & Slowly ramp up injected value for GNSS north position\\\cline{2-3}
                & \cellcolor[HTML]{dddddd}VBF & Gradually increase value added to GNSS north velocity\\\hline
\end{tabular}

    \end{table}

\begin{table*}[th!]
    \centering
    \footnotesize
        \begin{subtable}[t]{2.3in}
            \begin{tabular}{@{}|c|p{0.3in}|p{1.3in}|@{}}\hline
            \cellcolor[HTML]{bdbdbd}\textbf{MID}& \cellcolor[HTML]{bdbdbd}\textbf{Name} & \cellcolor[HTML]{bdbdbd}\textbf{Description}\\\hline\hline
            \iddef{M1}& Stealth $\chi^2$& Determines if the attack triggered the alarm under the default AD (\ie $\chi^2$) (\eq{\ref{eq:chisq}}).\\\hline
            \iddef{M2}& $\chi^2$ Max Stealthy $\epsilon$&  Maximum stealthy deviation achieved under  $\chi^2$ AD in meters.\\\hline
            \iddef{M3}& Stealth $\tau$&  Checks if $\tau$-AD (\eq{\ref{eq:threshold}}) raised an alarm.\\\hline
            \iddef{M4}& $\tau$ Max Stealthy $\epsilon$& Maximum stealthy deviation achieved under $\tau$-AD in meters\\\hline
        \end{tabular}
        \caption{Evaluation metrics}
        \label{tab:metrics}
        \end{subtable}
        \begin{subtable}[t]{4.5in}
                 \footnotesize
     \centering
     \begin{tabular}{@{}|@{}c@{}|c||c|c|c|c||c|c||c|c|@{}}\cline{3-10}
          \multicolumn{2}{c}{}&
          \multicolumn{4}{|c||}{\cellcolor[HTML]{ffffff}\textbf{Baseline}}&
          \multicolumn{2}{|c||}{\cellcolor[HTML]{bdbdbd}\textbf{Boiling Frog}}&
          \multicolumn{2}{|c|}{ \cellcolor[HTML]{FFD700}\textbf{\papername}}\\\hline
          \textbf{Mis.} & \textbf{MID} &
          \cellcolor[HTML]{ffffff}PCO&
          \cellcolor[HTML]{ffffff}PRO&
          \cellcolor[HTML]{ffffff}VCO&
          \cellcolor[HTML]{ffffff}VRO&
          \cellcolor[HTML]{dddddd}PBF&
          \cellcolor[HTML]{dddddd}VBF&
          \cellcolor[HTML]{FFFAA0}NC&
          \cellcolor[HTML]{FFFAA0}DB\\\hline\hline
          \multirow{4}{*}{\circleMission} & M1&
          \greencheck &
          \greencheck &
          \greencheck &
          \greencheck &
          \greencheck &
          \greencheck &
          \greencheck &
          \greencheck\\\cline{2-10}
          & M2  &
          1.097 &
          0.365 &
          2.730 &
          0.934 &
          1.122 &
          2.756 &
          \cellcolor[HTML]{CBC3E3} 110.62 & 23.78\\\cline{2-10}
          & M3 &
          \redxmark &
          \redxmark &
          \redxmark &
          \redxmark &
          (\greencheck/\redxmark) &
          \redxmark &
          (\greencheck/\redxmark)
          & \greencheck \\\cline{2-10}
          & M4 &
          \redxmark &
          \redxmark &
          \redxmark &
          \redxmark &
          0.111 &
          \redxmark &
          \cellcolor[HTML]{CBC3E3} 77.09 & 23.78\\\hline
          \hline
          \multirow{4}{*}{\linearMission} & M1 &
          \greencheck  &
          \greencheck  &
          \greencheck  &
          \greencheck  &
          \greencheck  &
          \greencheck  &
          \greencheck &
          \greencheck \\\cline{2-10}
          & M2 &
          1.030 &
          0.300 &
          2.745 &
          0.891 &
          0.356 &
          0.893 &
          3.56 &
          \cellcolor[HTML]{CBC3E3} 11.10 \\\cline{2-10}
          & M3 &
          \redxmark &
          \redxmark &
          \redxmark &
          \redxmark &
          (\greencheck/\redxmark) &
          \redxmark &
          \greencheck &
          \greencheck\\\cline{2-10}
          & M4 & \redxmark& \redxmark& \redxmark& \redxmark& 0.0936 &  \redxmark& 3.56 & \cellcolor[HTML]{CBC3E3} 11.10\\\hline
          \hline
          \multirow{4}{*}{\holdMission} & M1 &
          \greencheck  &
          \greencheck  &
          \greencheck  &
          \greencheck  &
          \greencheck  &
          \greencheck  &
          \greencheck &
          \greencheck \\\cline{2-10}
          & M2 &
          1.018 &
          0.357 &
          2.756 &
          1.135 &
          0.925 &
          2.497 &
          50.62& \cellcolor[HTML]{CBC3E3}289.26 \\\cline{2-10}
          & M3 &
          \redxmark &
          \redxmark &
          \redxmark &
          \redxmark &
          \redxmark &
          \redxmark &
          (\greencheck/\redxmark) &
          (\greencheck/\redxmark)\\\cline{2-10}
          & M4 &
          \redxmark &
          \redxmark &
          \redxmark &
          \redxmark &
          \redxmark &
          \redxmark &
          \cellcolor[HTML]{CBC3E3} 17.72& 9.47\\\hline
     \end{tabular}
            \caption{
                Attack evaluation for each mission (Mis.)
            }
            \label{tab:result:eval}
        \end{subtable}
    \caption["Short" caption without tikz code]{
        \label{tab:result}
        Attack evaluation on the three mission scenarios in simulation:
        circle (\circleMissionSmall ), linear (\linearMissionSmall ) and hold (\holdMissionSmall ).
        The attack that resulted with the most deviation given a metric is shaded with purple.
        \greencheck indicates that the attack remained stealthy throughout the mission in all trials, (\greencheck/\redxmark) indicates that attacked start out stealthy for all trials but became overt for at least one of the trials, and \redxmark indicates that the attack was not stealthy to begin with for at-least one of the trials.
    }
\end{table*}
    
    Each mission poses different challenges for the attacker.
    The circle mission results in the vehicle rotation while moving.
    The attack becomes complex as the effects of attitude and angular velocity must be considered.
    Linear mission represents a standard mission of going from point A to point B without rotation.
    An attacker must ensure that the vehicle thinks that it converged to the mission setpoints.
    The hold mission represents scenarios where there is little to no movement, making it harder for the attacker to exploit estimation error.

\psection{AD Parameter}
     We compare the attack performance on both default onboard AD ($\chi^2$) and the threshold based AD ($\tau$) with mission tuned thresholds to show that bypassing $\tau$ has stricter requirements.
     It highlights the significance of \papername{} by demonstrating that while \papername{} can bypass $\tau$, baseline methods cannot.
    
    $\chi^2$ AD has default thresholds of $\eta =5$.
    However for $\tau$ AD, the threshold range ($\tau_{min}, \tau_{max}$)  were chosen based on the nominal distribution of the residual values for each mission.
    Specifically, the standard deviation of the distribution was calculated and we removed outliers outside two standard deviations.
    Then the $2.5\%$ and $97.5\%$ percentile were selected as $\tau_{min}$ and $\tau_{max}$ respectively.

\label{sec:eval:subsec:sim}

\psection{Experiment Setup}
Experiments used the PX4 autopilot in a software-in-the-loop (SITL) simulation mode with Gazebo\cite{gazebo} plugin with the default environment (\ie no environmental hazards such as wind).
The simulation ran on quad-core intel i5, Ubuntu 22.04, 16GB RAM, AMD Radeon Vega 20 GPU.
During the course of simulation, the autopilot's state estimation module exfiltrates the input specified in \tab{\ref{tab:collected_data_type}}
whenever the sensor value is ready before the next  state update.
The module receives the spoof value after the input exfiltration and injects the value into the GNSS prior to $F$.

\psection{Process}
We apply each of the attack approaches (2 \papername{} and 6 baseline as shown in  \tab{\ref{tab:attack_types} and \ref{tab:naive_attacks}} respectively) against the three missions (\ie hold, linear, circle) each representing different mission kinematics.
Each experiment is repeated 10 times and evaluated using metrics from \S\ref{sec:eval:subsec:metrics}.
Results of the experiments are organized in \tab{\ref{tab:result}} which shows the average performance of the attacks for each mission measured using our metrics.

\subsection{\ref{rq:success} To what extent does \papername{} succeed?}
\label{sec:rq:success}

\begin{figure*}[h!]
    \def\plotWidth{0.24\linewidth}
    \centering
    \begin{minipage}{\linewidth}
    \def\legendPadding{2.5mm}
    \centering
    \footnotesize
    \begin{tabular}{|@{\hspace{\legendPadding}}c@{\hspace{\legendPadding}}c@{\hspace{\legendPadding}}c@{\hspace{\legendPadding}}c@{\hspace{\legendPadding}}c@{\hspace{\legendPadding}}c@{\hspace{\legendPadding}}c@{\hspace{\legendPadding}}c@{\hspace{\legendPadding}}|}\hline
        \begin{tikzpicture}[scale=1]
            \draw[line width=0.5mm, color=black] (0,0) -- (0.5,0);
        \end{tikzpicture}
        : Planned 
        &
        \begin{tikzpicture}[scale=1]
            \draw[line width=0.5mm, normal] (0,0) -- (0.5,0);
        \end{tikzpicture}
        : Nominal
        &
        \begin{tikzpicture}[scale=1]
            \draw[line width=0.5mm, dashed, color=op ] (0,0) -- (0.5,0);
        \end{tikzpicture}
        : System POV
        &
        \begin{tikzpicture}[scale=1]
            \draw[dashed, line width=0.5mm, normal] (0,0) -- (0.5,0);
        \end{tikzpicture}
        : $\tau$ Thresh
        &
        \begin{tikzpicture}[scale=1]
            \draw[dashed, line width=0.5mm, black] (0,0) -- (0.5,0);
        \end{tikzpicture}
        : $\chi^2$ Thresh
        &
        \begin{tikzpicture}[scale=1]
            \draw[line width=0.5mm, attack] (0,0) -- (0.5,0);
        \end{tikzpicture}
        : Stealthy
        &
                \begin{tikzpicture}
                    \path (0,0) pic[line width=0.3mm, rotate = 0] {cross=3pt};
                \end{tikzpicture}: Stealth Loss Point
                &
        \begin{tikzpicture}[scale=1]
            \draw[dashed, line width=0.5mm, red] (0,0) -- (0.5,0);
        \end{tikzpicture}
        : Overt
        \\\hline
    \end{tabular}
\end{minipage}
    \begin{minipage}{\plotWidth}
        \begin{subfigure}{\linewidth}
            \footnotesize
            \includegraphics[height=3.5cm, width=\linewidth]{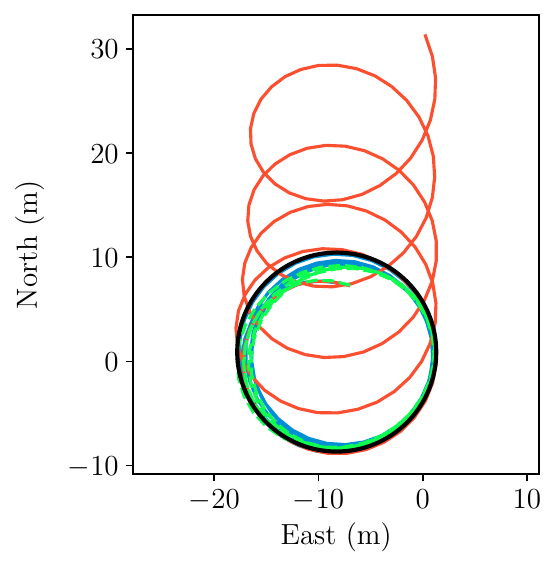}
            \caption{\textit{DB Mission Trajectory}}
            \label{fig:db_perf:sub:traj}
        \end{subfigure}
    \end{minipage}
    \begin{minipage}{\plotWidth}
        \begin{subfigure}{\linewidth}
            \includegraphics[height=3.5cm, width=\textwidth]{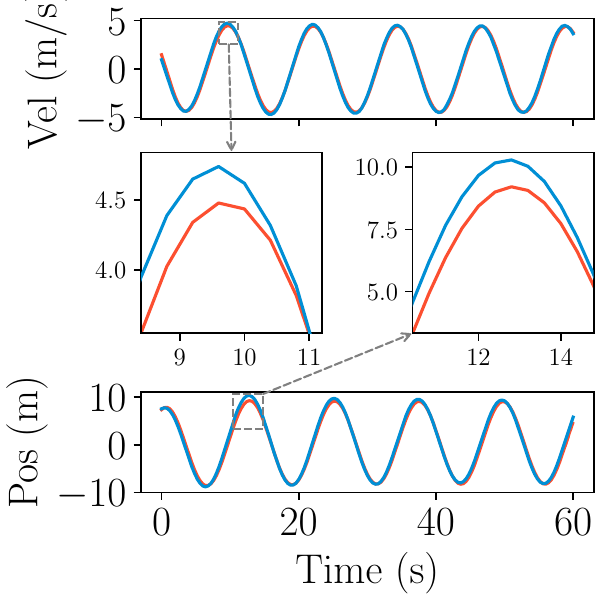}
            \caption{\textit{DB Observed GNSS History}}
            \label{fig:db_perf:sub:velpos}
        \end{subfigure}
    \end{minipage}
    \hspace{0.005\linewidth}
    \begin{minipage}{\plotWidth}
        \begin{subfigure}{\linewidth}
            \includegraphics[height=3.5cm, width=\textwidth]{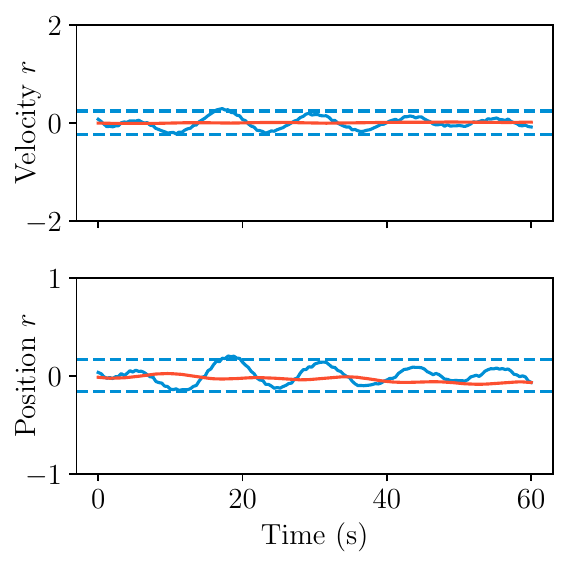}
            \caption{\textit{DB Residual History}}
            \label{fig:db_perf:sub:tau}
        \end{subfigure}
    \end{minipage}
    \begin{minipage}{\plotWidth}
        \begin{subfigure}{\linewidth}
            \includegraphics[height=3.5cm, width=\linewidth]{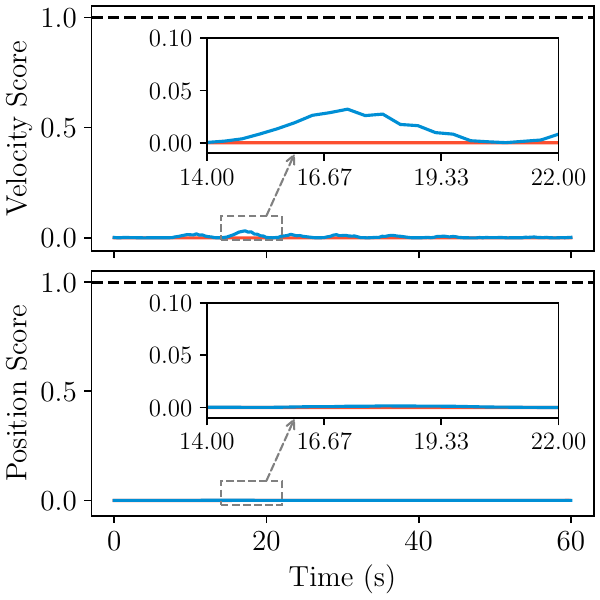}
            \caption{\textit{DB $\chi^2$ Score} 
            }
            \label{fig:db_perf:sub:chi}
        \end{subfigure}
    \end{minipage}
    \caption{
    \papername's directional bias (DB) attack on circle mission.
    \textit{(\subref{fig:db_perf:sub:traj})} shows that system POV trajectory (green) is not distinguishable from the nominal where as the attack trajectory, colored in red, deviated far away.
    During the attack, GNSS velocity and position readings are seemingly nominal as shown in  \textit{(\subref{fig:db_perf:sub:velpos})} where the difference is minuscule.
    \textit{(\subref{fig:db_perf:sub:tau})} Resulting residual stayed within the $\tau$ threshold, avoiding the activation of the alarm.
    Onboard detector's inefficacy is demonstrated in \textit{(\subref{fig:db_perf:sub:chi})} where $\chi^2$ score remained near-zero throughout.
    }
    \label{fig:db_perf}
    \vspace{-3mm}
\end{figure*}

\begin{figure}[h]
    \centering
    \begin{minipage}[t]{\columnwidth}
        \centering
            \begin{minipage}{\columnwidth}
        \centering
        \footnotesize
        \begin{tabular}{|lll|}\hline
            \begin{tikzpicture}[scale=1]
                \draw[line width=1.5mm, color=tbdNormal] (0,0) -- (0.5,0);
            \end{tikzpicture}
            : Nominal Traj Range 
            &
            \begin{tikzpicture}[scale=1]
                \draw[line width=1.5mm, color=tbdSysPOV ] (0,0) -- (0.5,0);
            \end{tikzpicture}
            : Sys POV 
            &
            \begin{tikzpicture}[scale=1]
                \draw[line width=1.5mm, color=tbdAttack] (0,0) -- (0.5,0);
            \end{tikzpicture}
            : Attack Traj
            \\\hline
        \end{tabular}
    \end{minipage}
        \includegraphics[height=5cm, width=\linewidth]{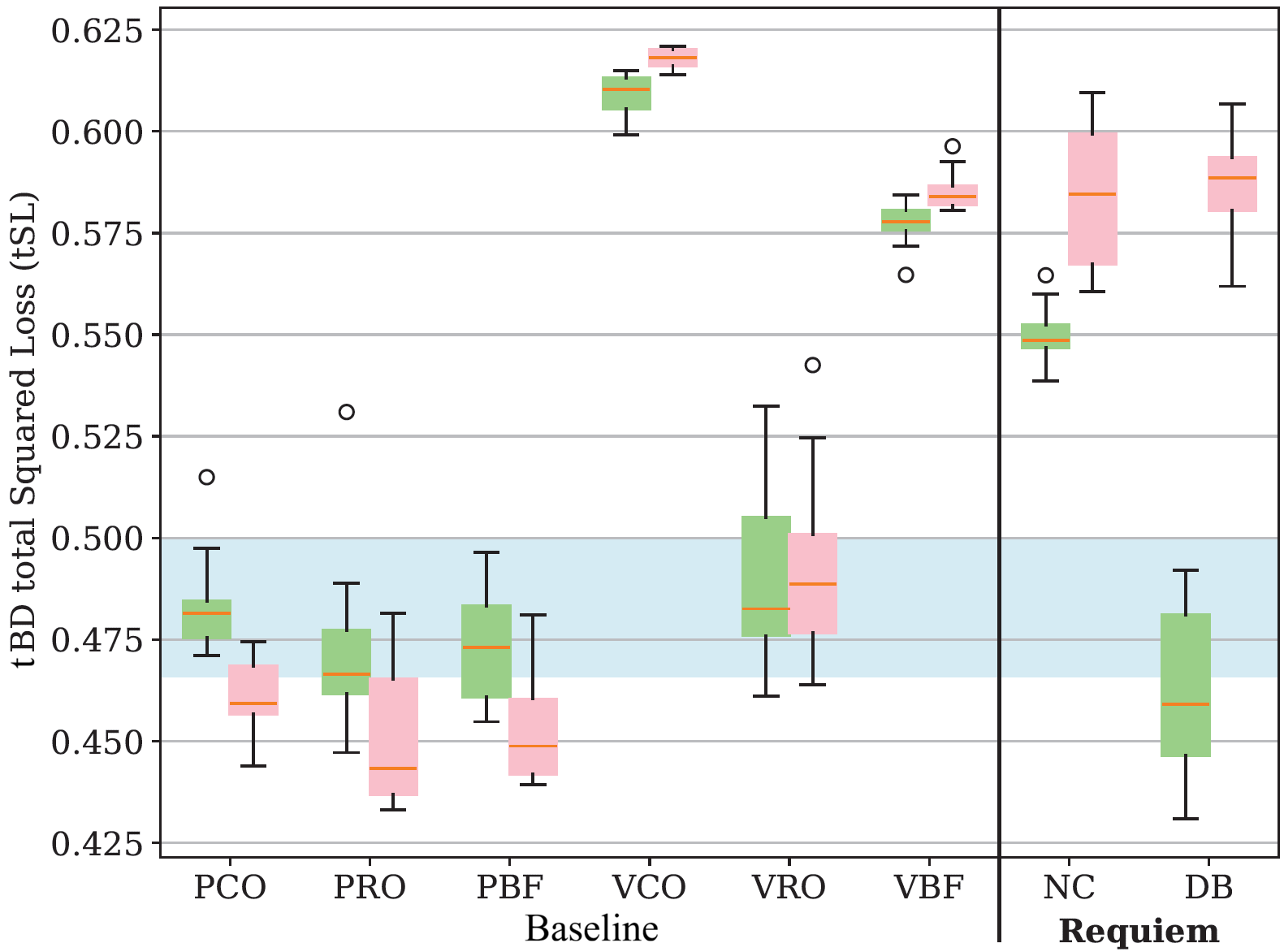}
        \caption{
            \label{fig:circle_tbd}
            Total Bregman Divergence (\TBD) between the planned, attack and the system POV trajectories.
            X-axis lists each attack and y-axis shows their corresponding score.
            A larger difference between the green and the pink box indicates a larger deviation.
            The plot shows that \papername{} attacks achieve significant deviation whereas baseline attacks fail to do so.
        }
    \end{minipage}
    \vspace{-\baselineskip}
\end{figure}

The objective of a stealthy attack is to avoid tripping any anomaly detectors and causing the system POV trajectory to seem nominal \wrt the planned mission while the {\em true trajectory} significantly deviates from the mission parameters due to the attacks (\ie a meaningful deviation).
Justification and definitions of `stealthiness' and `meaningfulness' are discussed in the subsequent section \S \ref{sec:rq:meaningful} and \S\ref{sec:rq:stealthy} respectively.
On both fronts, \papername \textit{successfully} achieved stealth since it {\em caused significant deviation without triggering anomaly detectors} especially for the circle and linear missions as seen in \tab{\ref{tab:result}}.
For the hold mission, \papername{} was successful until an average of 17.72m of deviation was achieved after which it was no longer stealthy.
The reason for the break in stealth is that the hold mission has a very limited errors in movement that \papername{} can exploit.
However all of the baseline attacks (\tab{\ref{tab:naive_attacks}}) failed to be stealthy for {\em all} missions, displaying the non-triviality of the problem.
In the rest of this section, we discuss the attack \wrt the circle mission {as it exhibits the most complex kinematics, hence harder to stealthily spoof, especially for a longer period of time}; the details and results from the other missions are on the website\footnote{\label{fn:section_6_website}\paperwebsite{}} due to the space constraints.

\begin{figure}[h!]
    \centering
        \begin{minipage}{\columnwidth}
            \centering
            \def\legendPadding{\hspace{2.5mm}}
            \footnotesize
            \begin{tabular}{|@{\legendPadding}l@{\legendPadding}l@{\legendPadding}l@{\legendPadding}l@{\legendPadding}|}\hline
                \begin{tikzpicture}[scale=1]
                    \draw[line width=0.5mm, color=black] (0,0) -- (0.5,0);
                \end{tikzpicture}
                : Planned 
                &
                \begin{tikzpicture}[scale=1]
                    \draw[line width=0.5mm, normal] (0,0) -- (0.5,0);
                \end{tikzpicture}
                :  Nominal
                &
                \begin{tikzpicture}[scale=1]
                    \draw[dashed, line width=0.5mm, color=op ] (0,0) -- (0.5,0);
                \end{tikzpicture}
                : Sys POV
                &
                \begin{tikzpicture}[scale=1]
                    \draw[dashed, line width=0.5mm, normal] (0,0.1) -- (0.5,0.1);
                \end{tikzpicture}
                : $\tau$ Thresh
                \\
                \begin{tikzpicture}[scale=1]
                        \draw[line width=0.5mm, attack] (0,0) -- (0.5,0);
                        \end{tikzpicture}
                        : Stealthy
                         &
                \multicolumn{2}{c}{
                    \begin{tikzpicture}
                        \path (0,0) pic[line width=0.3mm, rotate = 0] {cross=3pt};
                    \end{tikzpicture}: Stealth Loss Point
                }
                &
                \begin{tikzpicture} [scale=1]
                            \draw[dashed, line width=0.5mm, attack] (0,0) -- (0.5,0);
                        \end{tikzpicture}
                        : Overt
                \\\hline
            \end{tabular}
        \end{minipage}
    \begin{minipage}{0.49\columnwidth}
        \centering
        \begin{subfigure}{\linewidth}
            \includegraphics[height=3.5cm, width=\linewidth]{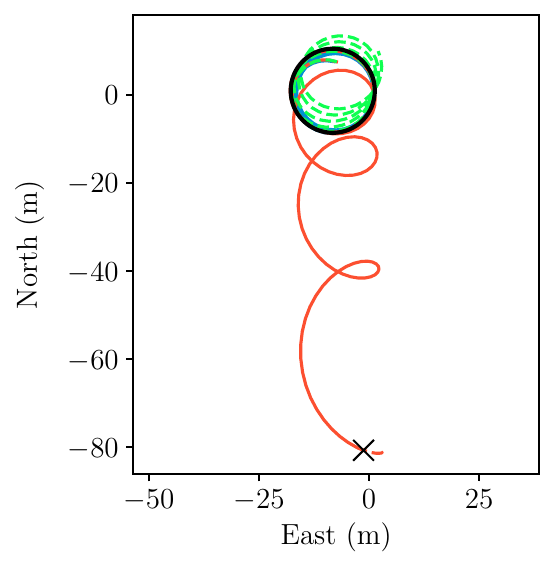}
            \caption{\textit{NC Mission Trajectory}}
            \label{fig:nc_perf:sub:traj}
        \end{subfigure}
    \end{minipage}
    \begin{minipage}{0.49\columnwidth}
        \centering
        \begin{subfigure}{\linewidth}
            \includegraphics[height=3.5cm, width=\linewidth]{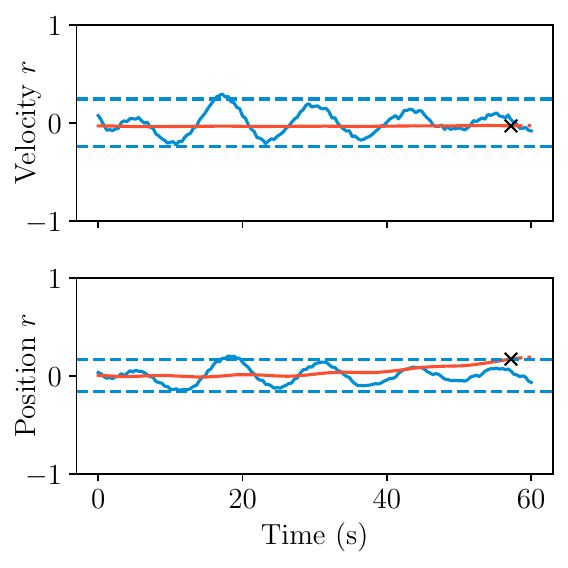}
            \caption{
            \textit{NC Residual}
            }
            \label{fig:nc_perf:sub:tau}
        \end{subfigure}
    \end{minipage}
    \caption{\papername's no correction (NC) attack on circle mission. (\subref{fig:nc_perf:sub:traj}) shows the UAV deviating a significant amount but became overt near the end. (\subref{fig:nc_perf:sub:tau}) shows the point of the stealth loss.
    The loss is due to receiving large position values (\ie >77m) as input which is far outside of the training distribution.
    }
    \label{fig:nc_perf}
    \vspace{-\baselineskip}
\end{figure}

\papername{} was indeed stealthy since it did not trigger any alarms --- for both anomaly detectors --- for the duration of the mission and the system POV was nominal\footnote{we use `nominal' and `expected' interchangeably}.
\fig{\ref{fig:db_perf}} shows:
(\subref{fig:db_perf:sub:traj}) trajectory (east-north coordinates),
(\subref{fig:db_perf:sub:velpos}) observed GNSS sensor (time vs. velocity/position),
(\subref{fig:db_perf:sub:tau}) residual history (time vs. velocity/position residuals), and
(\subref{fig:db_perf:sub:chi}) $\chi^2$ score (time vs. velocity/position scores).
The performance of \papername{}'s direction bias (DB) attack in \fig{\ref{fig:db_perf:sub:tau}} shows that the {\em residual stayed within the threshold range for the duration of the mission}.
Notice that in \fig{\ref{fig:db_perf:sub:traj}}, the system POV trajectory is near exact to the nominal trajectory.
Consequently in \fig{\ref{fig:circle_tbd}}, the total Bregman divergence (\TBD{}) value of the system POV trajectory is within the nominal trajectory range.

    We can see that although the attack residual remained stealthy for DB attack illustrated in \fig{\ref{fig:db_perf:sub:tau}}, the residual is slowly reaching the lower threshold as the deviation increases.
    The effect is due to the larger discrepancy between the estimated state and the true sensor value when compared to those from the data collection stage (\ie training set).
    The effect is more prominent in \papername{}'s NC attack as shown in \fig{\ref{fig:nc_perf}} where the attack exploits the downward momentum, causing faster deviation at the cost of becoming \textit{overt} (around $70m$ of deviation).
    For DB, had the vehicle kept deviating away from the planned trajectory for a longer time, the attack would eventually become \textit{overt}.
    However, {\em the fact that spoofer still remained stealthy even after about $25m$ of deviation, far exceeding the scenario captured in the training set, indicates that the model generalized well}.

\begin{figure}[h]
    \centering
        \begin{minipage}{\columnwidth}
            \centering
            \def\legendPadding{\hspace{2.5mm}}
            \footnotesize
            \begin{tabular}{|@{\legendPadding}l@{\legendPadding}l@{\legendPadding}l@{\legendPadding}l@{\legendPadding}|}\hline
                \begin{tikzpicture}[scale=1]
                    \draw[line width=0.5mm, color=black] (0,0) -- (0.5,0);
                \end{tikzpicture}
                : Planned 
                &
                \begin{tikzpicture}[scale=1]
                    \draw[line width=0.5mm, normal] (0,0) -- (0.5,0);
                \end{tikzpicture}
                :  Nominal
                &
                \begin{tikzpicture}[scale=1]
                    \draw[dashed, line width=0.5mm, color=op ] (0,0) -- (0.5,0);
                \end{tikzpicture}
                : Sys POV
                &
                \begin{tikzpicture}[scale=1]
                    \draw[dashed, line width=0.5mm, normal] (0,0.1) -- (0.5,0.1);
                \end{tikzpicture}
                : $\tau$ Thresh
                \\
                \begin{tikzpicture}[scale=1]
                        \draw[line width=0.5mm, attack] (0,0) -- (0.5,0);
                        \end{tikzpicture}
                        : Stealthy
                         &
                \multicolumn{2}{c}{
                    \begin{tikzpicture}
                        \path (0,0) pic[line width=0.3mm, rotate = 0] {cross=3pt};
                    \end{tikzpicture}: Stealth Loss Point
                }
                &
                \begin{tikzpicture} [scale=1]
                            \draw[dashed, line width=0.5mm, attack] (0,0) -- (0.5,0);
                        \end{tikzpicture}
                        : Overt
                \\\hline
            \end{tabular}
        \end{minipage}
    \begin{minipage}{0.49\columnwidth}
        \begin{subfigure}{\linewidth}
            \includegraphics[height=3.5cm, width=\linewidth]{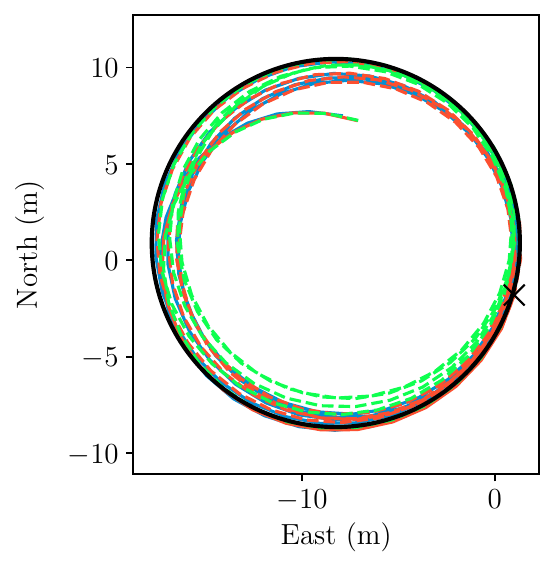}
            \caption{PBF Mission Trajectory}
            \label{fig:pbf_perf:sub:traj}
        \end{subfigure}
    \end{minipage}
    \begin{minipage}{0.49\columnwidth}
        \begin{subfigure}{\linewidth}
            \includegraphics[height=3.5cm, width=\linewidth]{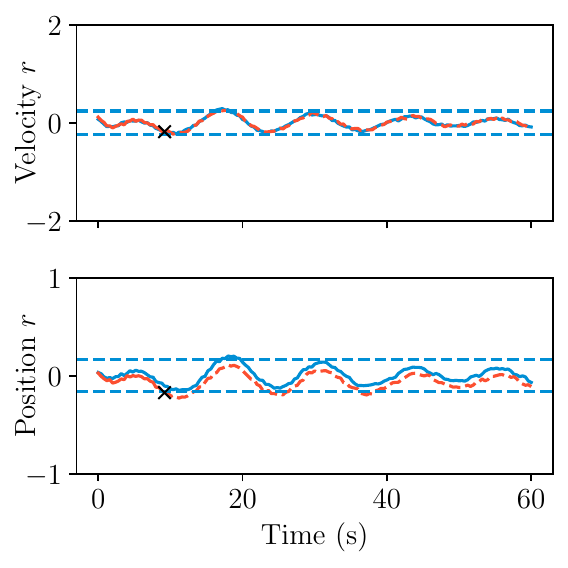}
            \caption{PBF Residual}
            \label{fig:pbf_perf:sub:tau} 
        \end{subfigure}
    \end{minipage}
    \caption{Position boiling frog (PBF) attack on circle mission.
    (\subref{fig:pbf_perf:sub:traj}) shows negligible effect of PBF on the trajectory but becomes overt after traveling half circle.
    (\subref{fig:pbf_perf:sub:tau}) shows the position residual gradually straying away from the nominal residual resulting in an alarm.
    }
    \label{fig:pbf_perf}
\end{figure}

\papername{} achieved a 23m northward deviation for a helical trajectory, significantly different from the intended circle as shown in \fig{\ref{fig:db_perf:sub:traj}}.
As a result, the \TBD{} value of the attack trajectory (in \fig{\ref{fig:circle_tbd}}) is significantly higher than the system POV.
In contrast, position boiling-frog attack (PBF), the only
baseline method that had partial success (shown in \fig{\ref{fig:pbf_perf}}
) achieved maximum of about $0.111m$ \textit{stealthy} deviation, twice the average nominal error as shown in \tab{\ref{tab:simulation_missions}}.
This is not comparable to \papername{} which is {\em two orders of magnitude higher}.
Even at the cost of becoming \textit{overt} and reaching about $1m$ of deviation, the PBF's attack trajectory is similar to the nominal trajectory, thus not causing any real deviations.

The effect of the loss function on the spoofer is noticeable as the Direction Bias successfully caused the trajectory to shift northward.
Notice at the start of the circle mission, there is a counter clockwise momentum causing the vehicle to head south (visible in \fig{\ref{fig:pbf_perf:sub:traj}}).
{Therefore, contrast between NC (\fig{\ref{fig:nc_perf}}) and DB (\fig{\ref{fig:db_perf:sub:traj}}) indicates that DB successfully biases deviation toward the attacker's chosen direction while NC follows the vehicle's existing momentum.}

\subsection{\ref{rq:realworld}
\label{sec:rq:realworld} How well does \papername{} work on real-world?}
\begin{figure}[t]
    \centering
    \begin{minipage}{2.2in}
        \begin{subfigure}{2.2in}
            \centering
            \includegraphics[width=1.6in]{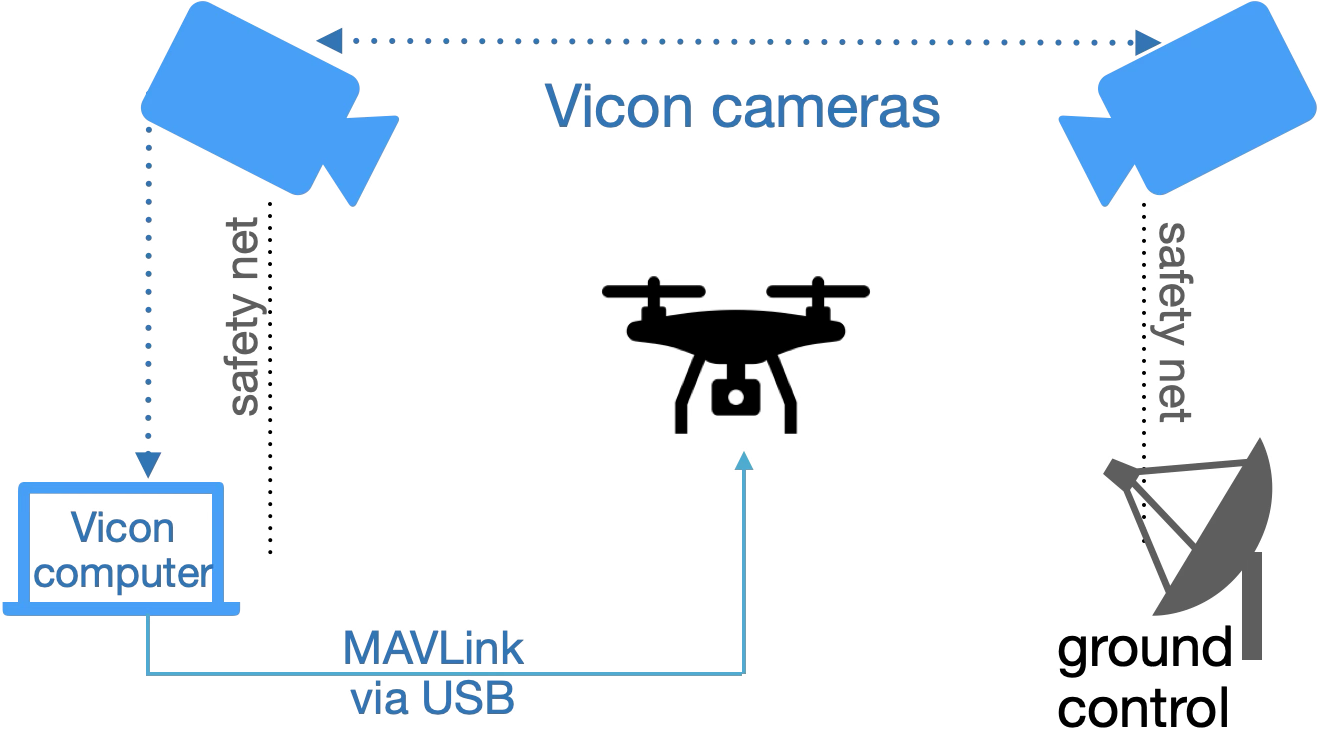}
            \subcaption{Setup Diagram}
            \label{fig:realworld_setup:diagram} 
        \end{subfigure}
    \end{minipage}
    \begin{minipage}{1.1in}
        \begin{subfigure}{1.1in}
            \includegraphics[width=1.1in]{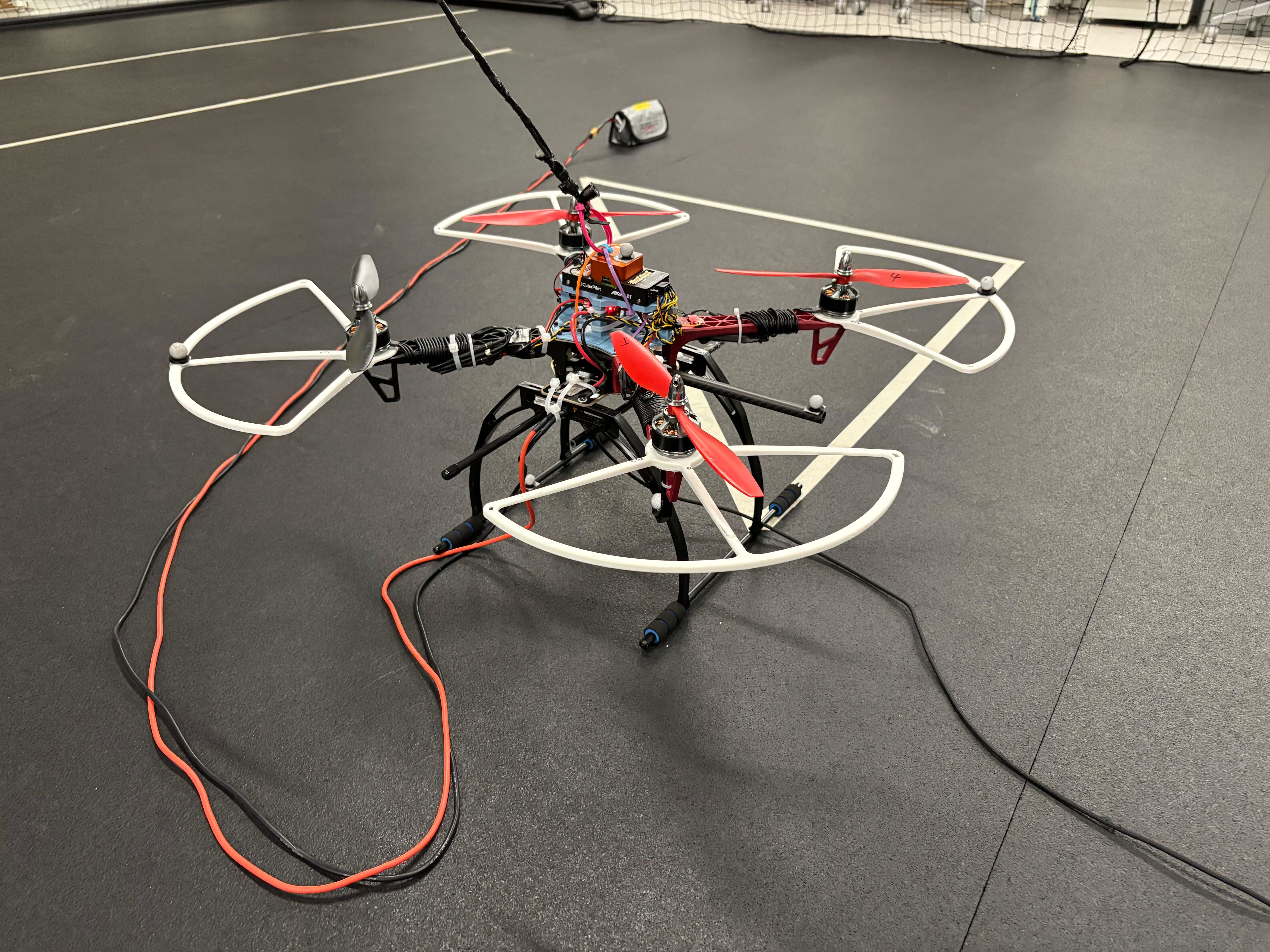}
            \subcaption{ Quadrotor}
            \label{fig:realworld_hardware}
        \end{subfigure}
    \end{minipage}
    \caption{Real-world experiment setup.
            (\subref{fig:realworld_setup:diagram})
            Vicon motion capture system sends data via MAVLink over USB for reliable reception.
            (\subref{fig:realworld_hardware}) shows the hardware implementation.
        }
    \label{fig:realworld_setup}
    \vspace{-1em}
\end{figure}

We extend our simulation results to the real-world to demonstrate practicality of the attack on physical drones.
Due to space constraints, we show the simplest example here (\ie linear mission) for the real-world evaluation while the rest are uploaded to the website\footnoteref{fn:section_6_website}.
To show generalizability, we attack a different target function (\ie sensor fusion of motion capture data).
To demonstrate practicality, the attacker does not have access to a query server as specified in \S\ref{sec:req:imp}.
Instead, \papername{} was trained on approximately 1 hour of random walk mission (155 flights), yielding 94543 training examples.

\psection{Transfering Simulation Results to the Real-World}
To demonstrate that simulation results are transferable to the real-world, we configured the quadrotor parameters to match the simulation quadrotor as closely as possible (using a single IMU).
Since indoor environments attenuate GNSS\footnote{We are limited to indoor experiments since we are \textit{restricted from flying drones outside by local laws}. We have a \textit{fully functional UAV lab} used for our experiments.} and introduce magnetometer/barometer interference, we use a Vicon motion capture system \cite{vicon_valkyrie} for position, heading and altitude --- effectively replacing GNSS, magnetometer and barometer.
Vicon publishes at $10Hz$ with centimeter accuracy, \textit{making it harder to attack than GNSS, especially in a stealthy manner} ($5Hz$, higher noise).
Indoor testing also eliminates weather as a confounding variable.

\begin{figure}
    \centering
    \begin{minipage}{\columnwidth}
        \includegraphics[width=\textwidth]{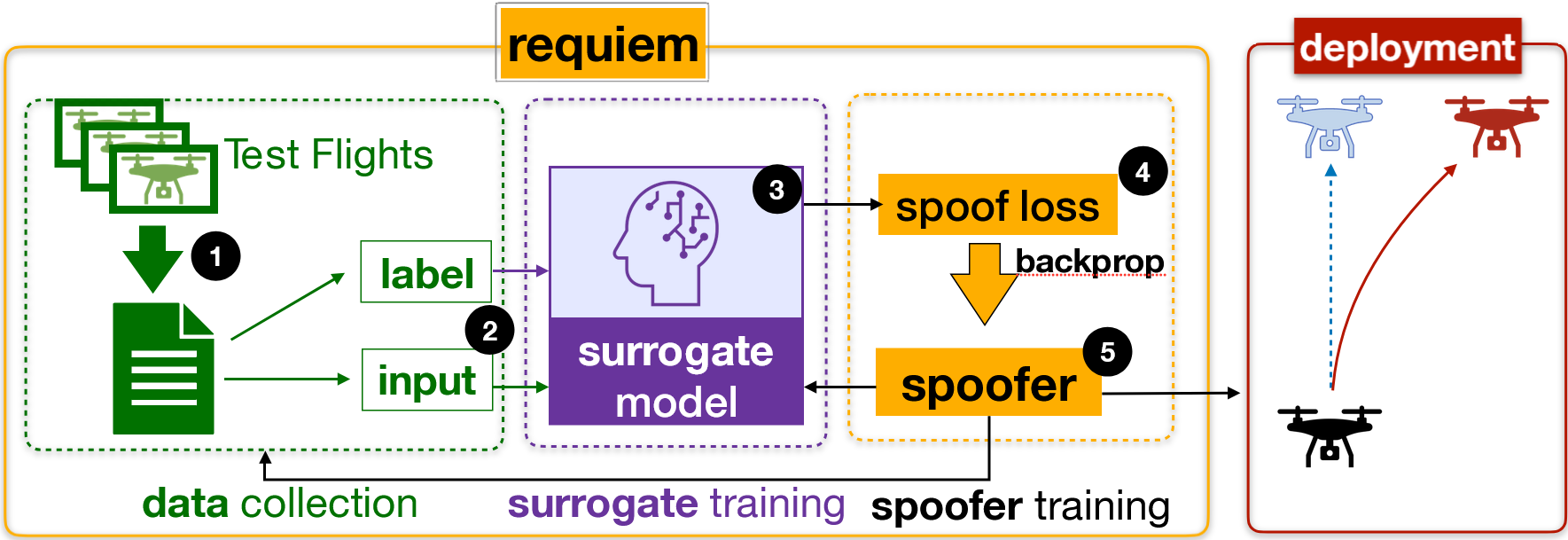}
        \caption{\papername{} Hardware-only Pipeline.
        {%
        While simulation provides a convenient means of collecting training data, it may not always be possible to create a SITL or query constantly due to software/policy restrictions.
        In such scenarios, \papername{} can instead be trained using \textit{data collected directly from a physical vehicle}, of the same type as the target.}
        Keep in mind, these preliminary flights {\em differ} from the actual deployment missions.
        \circledReq{1} Multiple preliminary flights (different from deployment missions) are performed to
        \circledReq{2} collect target function I/O and
        \circledReq{3} train a surrogate model.
        \circledReq{4} A spoofer is trained against it then 
        \circledReq{5} tested in the preliminary flights, generating new data and restarting the cycle.
        Once satisfactory performance is achieved, the spoofer is deployed.
        Overall, more data is collected compared to the simulation pipeline.
        }
        \label{fig:hardware_pipeline}
    \end{minipage}
\end{figure}

\psection{\papername{} Setup}
We did not use simulation to collect data in the hardware evaluation.
Instead, \papername{} followed the pipeline as shown in \fig{\ref{fig:hardware_pipeline}}.
Instead of input exploration in \fig{\ref{fig:main_pipeline}}, we collect data on the performance of the preliminary spoofer (\ie loop back step in \circledReq{5}).
This approach covers the scenarios of how the target function behaves when the spoofing occurs, refining the surrogate and spoofer training process.

\psection{Attack Implementation}
\fig{\ref{fig:realworld_hardware}} shows the quadrotor: four kv780 motors, CubeOrange+ flight controller running PX4 v1.15.4 on NuttX operating system.
It has STM32H753VI CPU (dual-core 400MHz) with 1MB of RAM.
The flying arena is bounded by a safety net.
When the quadrotor is close to the net (\ie within approximately 2m from the net), it is killed for safety.
Due to space constraints of the flying arena and safety considerations, we only evaluate linear and hold mission;
circular movements are difficult to execute safely as it is challenging to predict whether the quadrotor will collide with the net\footnote{We did run nominal physical circular mission for reference --- see the website \footnoteref{fn:section_6_website}.}.
The missions are scaled down to fit within the arena (as detailed in \app{\ref{appendix:missions}}).

Unlike the simulations, the attack inference runs in the \texttt{ekf2} module using the existing PX4 matrix library.
We used branchless ReLU (\ie bitshift to achieve the same effect) to avoid branch prediction penalties. 
The inference executes when the next Vicon sensor value is popped from the buffer.

We managed to make the spoofer a 3 layer DNN, resulting in 21821 bytes of model parameters that are loaded from an SD card during initialization. 
We verified that running the missions with the additional computation overhead due to the spoofer model inference (including a DNN of size equivalent to the spoofer used in the simulation) does not affect the quality of the flight --- and sufficient memory is available to hold the model parameters.

\psection{Experiment Setup}
The experiment is performed indoors with the setup shown in \fig{\ref{fig:realworld_setup}}.
The position and orientation information is provided to the quadrotor (as $\texttt{VISION\_POSITION\_ESTIMATE}$ MAVLink packets) using the motion capture camera, Vicon Valkyrie~\cite{vicon_valkyrie}.
Therefore, instead of attacking GNSS, we attack Vicon's north position value.
This also demonstrates how \papername{} can be easily adapted to target other types of sensors.
We evaluate \papername{}'s real-world NC attack and compare it against the corresponding simulation results.

\begin{figure}
    \centering
    \begin{minipage}{0.49\columnwidth}
        \begin{subfigure}{\linewidth}
            \includegraphics[height=3.5cm, width=\linewidth]{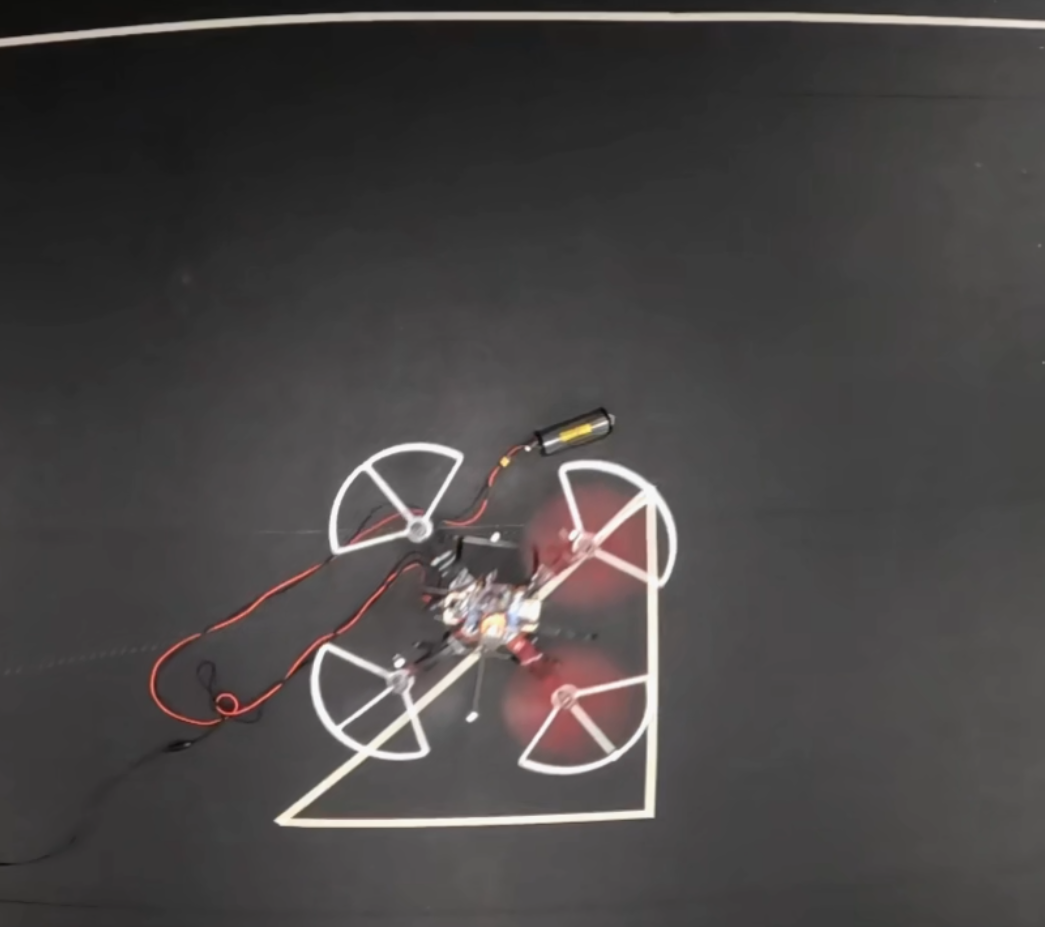}
            \caption{
                Nominal
                }
            \label{fig:realworld_experiment:demo:nominal}
        \end{subfigure}
    \end{minipage}
    \begin{minipage}{0.49\columnwidth}
        \begin{subfigure}{\linewidth}
            \includegraphics[height=3.5cm, width=\linewidth]{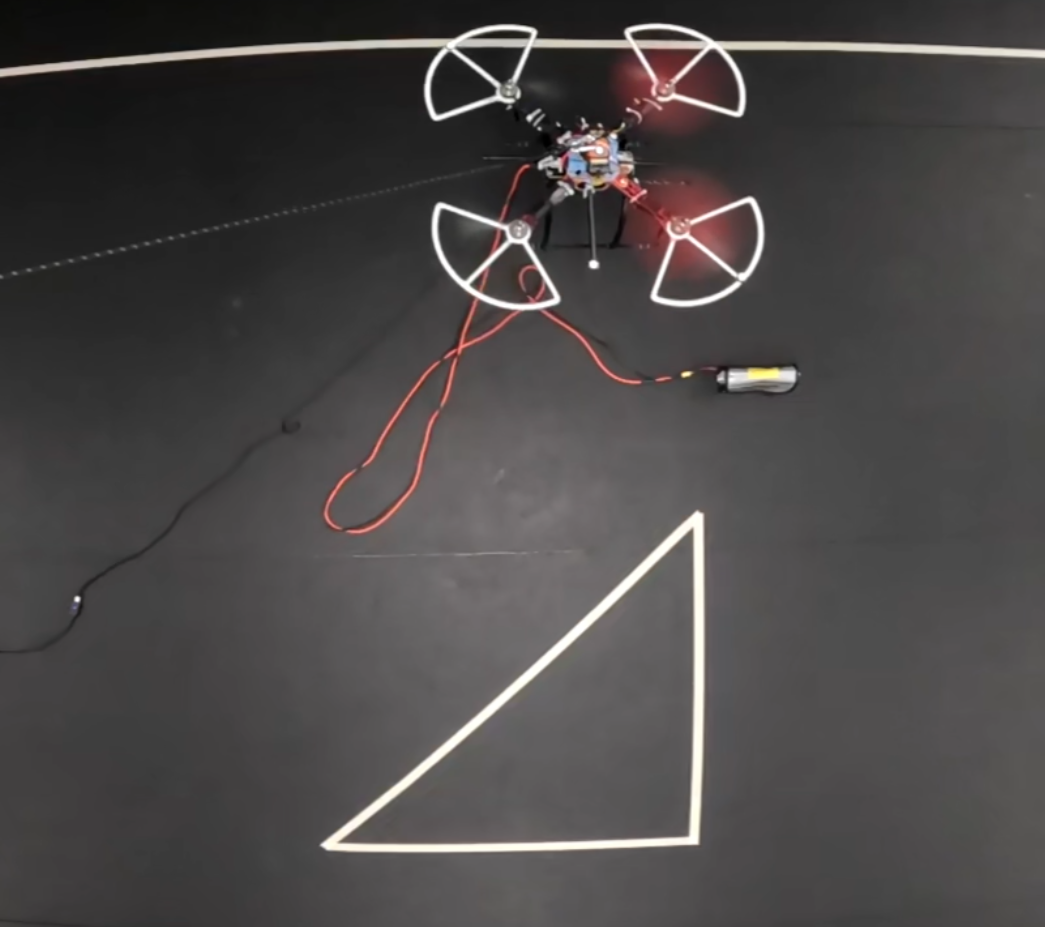}
            \caption{
                \papername{}
                }
            \label{fig:realworld_experiment:demo:requiem}
        \end{subfigure}
    \end{minipage}
    \caption{
        \label{fig:realworld_experiment:demo}
        Photos from the video of the physical experiment.
        A triangle is taped on the ground to indicate 1m east and 1m north from the takeoff point.
        Nominally, (\subref{fig:realworld_experiment:demo:nominal}) the quadrotor follows the mission trajectory, staying close to the triangle.
        However with \papername{}, (\subref{fig:realworld_experiment:demo:requiem}) it deviates towards north reaching close to the safety net.
    }
\end{figure}

\begin{figure}
    \centering
        \begin{minipage}{\columnwidth}
            \centering
            \def\legendPadding{\hspace{7.5mm}}
            \footnotesize
            \begin{tabular}{|@{\legendPadding}l@{\legendPadding}l@{\legendPadding}l@{\legendPadding}|}\hline
                \begin{tikzpicture}[scale=1]
                    \draw[line width=0.5mm, normal] (0,0) -- (0.5,0);
                \end{tikzpicture}
                : Nominal 
                &
                \begin{tikzpicture}[scale=1]
                    \draw[dashed, line width=0.5mm, color=op ] (0,0) -- (0.5,0);
                \end{tikzpicture}
                : System POV
                &
                \begin{tikzpicture}[scale=1]
                        \draw[line width=0.5mm, attack] (0,0) -- (0.5,0);
                        \end{tikzpicture}
                        : Stealthy
                \\\hline
            \end{tabular}
        \end{minipage}

    \begin{minipage}{0.49\columnwidth}
        \centering
        \begin{subfigure}{\linewidth}
            \includegraphics[height=3.5cm, width=\linewidth]{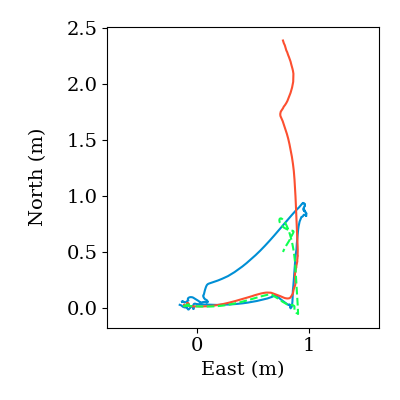}
            \caption{\label{fig:realworld:linear}\textit{Real-world evaluation}}
        \end{subfigure}
    \end{minipage}
    \begin{minipage}{0.49\columnwidth}
        \centering
        \begin{subfigure}{\linewidth}
            \includegraphics[height=3.5cm, width=\linewidth]{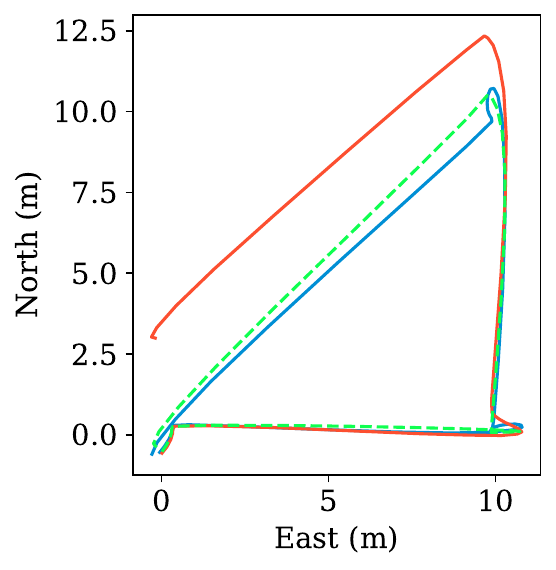}
            \caption{\label{fig:realworld:sim_linear}
            \textit{Simulation evaluation}
            }
        \end{subfigure}
    \end{minipage}
    \caption{
        \label{fig:realworld:linear_comparison}
        Comparison between real-world and simulation results for linear mission.
        Notice that in both cases, the SystemPOV trajectory (green) stays close to the nominal trajectory (blue) while the attack trajectory (red) deviates significantly.
        In the real-world, the plot only shows the trajectory until the quadrotor reaches the safety net.
    }
\end{figure}

\psection{Result}
In both linear (\fig{\ref{fig:realworld_experiment:demo}}) and hold \footnoteref{fn:section_6_website} missions, \papername{} caused significant deviations (\ie reached the net after about 1m of deviation) that we had to kill the quadrotor before it was able to finish the mission, thus demonstrating the attack's effectiveness in real-world.
We plot the trajectory in \fig{\ref{fig:realworld:linear_comparison}} which shows that the result of the attack in real-world is similar to the simulation result; the attack exploits the northward momentum to gain an increase in deviation.
This indicates that \textit{simulation results transfer to the real-world when the configurations closely mirror each other}.

\subsection{\ref{rq:weather} \label{sec:rq:weather} How do environmental factors affect the attacks?}

Environment factors (such as wind and other weather conditions) can cause the UAV to stray slightly away from its planned paths as shown in \fig{\ref{fig:windy_north_comp}} in \app{\ref{app:ref_scenario}}.
However, it can also be unpredictable and affect the mission performance of the UAV.
To test the robustness of \papername{}, we tested the spoofer, trained under the default environment (\ie no wind), under a windy weather scenario going steadily at 4m/s north.
We expected \papername{} to exploit the wind due to the natural momentum it provides to the UAV and deviate \textit{along} the direction of the wind.
However, the opposite was true:

\insight{
The EKF attempts to `fight' the wind to maintain position.
By suppressing the sensor data that reports this wind-induced drift, \papername{} tricks the controller into `over-correcting' against a non-existent force, effectively using the drone's own stabilization logic to drive it off course.
}

The result makes sense because the control already accounts for the wind during the takeoff (\ie before the mission start and the attack).
{Additional details are in \app{\ref{app:ref_scenario}}.}

\subsection{\ref{rq:learnable}
\label{sec:rq:learn} How well did surrogate models \textit{"learn"} a target function?}
A standard process for evaluating a model's performance in machine learning is by testing it against data that was not in the training set.
For this paper, the test set is the 8 separate mission runs addressed in \S\ref{sec:req:dc} and the performance measurement is the mean squared error loss.
It provides the following insight:

\insight{The surrogate models managed to learn the target function since important functions in safety critical systems, such as state estimator, are designed to be predictable, especially \wrt the correlation between their ins and outs.}

\noindent Please note the discussion of \textit{"learnability"} of the state estimation algorithm in \S\ref{sec:dis:learn}, reinforcing why surrogate models can \textit{"learn"} for many safety critical systems including autonomous vehicles.

\subsection{\papername{} State-of-the-Art Anomaly Detector (SAVIOR)}
\label{sec:eval:savior}

\begin{figure}
    \centering
        \begin{minipage}{\columnwidth}
            \centering
            \def\legendPadding{\hspace{2.5mm}}
            \footnotesize
            \begin{tabular}{|@{\legendPadding}l@{\legendPadding}l@{\legendPadding}l@{\legendPadding}|}\hline
                \begin{tikzpicture}[scale=1]
                    \draw[line width=0.5mm, normal] (0,0) -- (0.5,0);
                \end{tikzpicture}
                : Nominal
                &
                \begin{tikzpicture}[scale=1]
                        \draw[line width=0.5mm, attack] (0,0) -- (0.5,0);
                        \end{tikzpicture}
                        : Stealthy
                &
                    \begin{tikzpicture}
                        \path (0,0) pic[line width=0.3mm, rotate = 0] {cross=3pt};
                    \end{tikzpicture}: Stealth Loss Point
                \\
                \begin{tikzpicture}[scale=1]
                    \draw[dashed, line width=0.5mm, black] (0,0) -- (0.5,0);
                \end{tikzpicture}
                : $\tau$ Threshold
                &
                \begin{tikzpicture}[scale=1]
                    \draw[dashed, line width=0.5mm, color=attack ] (0,0) -- (0.5,0);
                \end{tikzpicture}
                : Overt
                &
                \\\hline
            \end{tabular}
        \end{minipage}
    \begin{minipage}{0.49\columnwidth}
        \centering
        \begin{subfigure}{\linewidth}
            \includegraphics[height=3.5cm, width=\linewidth]{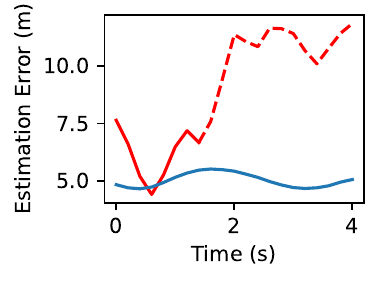}
            \caption{\label{fig:savior_est_error}\textit{SAVIOR Estimation Error}}
        \end{subfigure}
    \end{minipage}
    \begin{minipage}{0.49\columnwidth}
        \centering
        \begin{subfigure}{\linewidth}
            \includegraphics[height=3.5cm, width=\linewidth]{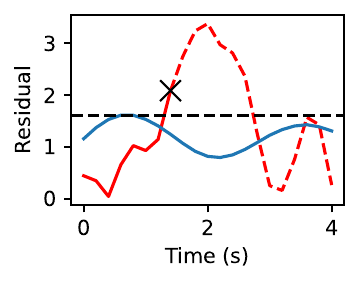}
            \caption{\label{fig:savior_residual}
            \textit{SAVIOR Residual}
            }
        \end{subfigure}
    \end{minipage}
    \caption{
        \label{fig:savior} Both plots show the attack performance against SAVIOR under the modified hold mission
        due to brittleness of SAVIOR.
        Even with the limitations, \papername{} show a promise of success as it shows higher estimation error than nominal with residual lower than nominal before the 1.5s mark.
    }
\end{figure}

SAVIOR~\cite{quinonez2020savior} is an EKF-based state-of-the-art anomaly detection method that is designed to detect sensor spoofing through improved modeling of vehicle's dynamics.
We found that SAVIOR's implementation was specific to a discontinued UAV, Intel Drone~\cite{intelDrone}, which is no-longer supported by PX4 since 2020.
We gave our best effort to translate the SAVIOR implementation to a more recent PX4 by only updating the PX4 module API but preserving the internal logic.
While the available research implementation of SAVIOR exhibited numerical instability 
\footnote{SAVIOR outputs high residual values before the 10m takeoff completes and the module, along with PX4, core dumps.}
in modern PX4 environments, \papername{} successfully identified adversarial inputs before these instability thresholds were reached. This demonstrates \papername{}'s robustness in finding injection values for stealthy attack.
\footnote{We are aware that SAVIOR was tuned for Intel Drone.}

We made adjustments to the mission parameters to reduce the chance of the SAVIOR module from breaking down. \papername{} still succeeded in spoofing SAVIOR as shown in \fig{\ref{fig:savior}}: we achieved up to 2.5m more estimation error than nominal with an indication of the residual being minimized.
Had the SAVIOR implementation been more stable, \papername{} would have further optimize against it.

\psection{Problem Formulation}
We define a stealthy attack against SAVIOR in context of \eq{\ref{eq:state_est}} as shown in \tab{\ref{tab:savior_problem_formulation}}.%
The target function $F$ is the $EKF$ function in SAVIOR with the static variables in the function (\ie state and error covariance matrix) treated as an inputs.
SAVIOR was tested on 1m hold mission using our simulation setup (\S\ref{sec:eval:subsec:sim}).

\psection{Adjustments}
We found that the SAVIOR system was sensitive to movement, therefore, we restricted the range of motion in the mission.
The parameter for random walk for data collection was set to 1m takeoff with average of 0.13m lateral difference between each random setpoints.
Even under such restrictions, SAVIOR raises alarms during the (nominal, not under attack) takeoff but the system did not crash.
In exchange for stability, the training set data distribution is narrowed, leading to potential overfitting.
Therefore, due to the narrowed residual distribution we use $\tau$-AD
\footnote{More restrictive than CUSUM used in SAVIOR as explained in \S \ref{sec:threatmodel:ad}}
with $\tau$ set to the max residual observed in a mission.

SAVIOR's EKF function is not a pure function since it has state and error covariance matrix static variables.
Therefore, to make the EKF function a pure function, we modified the EKF API so that only the static variables can be extracted, resulting in \tab{\ref{tab:savior_problem_formulation}} a vector of length 170.
Our approach is still opaque \wrt the SAVIOR's implementation of EKF since we do not extract any information about internal logic of the EKF and \papername{} only sees the inputs and the outputs.
We collected 20 missions where first 10 were used to train the surrogate and last 9 were used to train the spoofer, the remaining one served as validation for both (all for 50 epochs).
Due to narrower distribution (\S\ref{sec:eval:savior}), partitioning was designed to demonstrate cases where the surrogate observes an unseen data during the training of the spoofer.

Since we only performed NC attack against SAVIOR, we trained a surrogate model with linear activation, 10 layers of hidden layers where each layer has 50 nodes, and outputs north and east GNSS residuals.
The training of the surrogate resulted in near zero training and validation loss whereas for the spoofer the lowest train loss was 0.12 and 0.08 for validation loss.

\begin{table}[]
        \footnotesize
        \caption{\label{tab:savior_problem_formulation} Problem formulation for SAVIOR}
        \vspace{-2mm}
        \begin{tabular}{|p{0.5in}|>{\raggedright\arraybackslash}p{0.75in}|p{1.75in}|}
        \hline
        \cellcolor[HTML]{bdbdbd}\textbf{Type}          & \cellcolor[HTML]{bdbdbd}\textbf{Name}          & \cellcolor[HTML]{bdbdbd}\textbf{Description} \\ \hline
        Target Func. ($F$) & \texttt{EKF} & SAVIOR's implementation of \texttt{EKF}\\\hline
        \multirow{7}{*}{\parbox{1.27cm}{State\newline($\hat{x}_t$)\newline($\hat{x}_{t|t-1}$)} }
                                & Attitude ($\phi$)& (rad) in Roll, Pitch, Yaw frame\\ 
                                \cline{2-3} 
                                & Ang. Vel. ($\omega$)& Angular velocity measured in (rad/s)\\
                                \cline{2-3} 
                               & Position ($p$) and Velocity ($v$) & Represented in (m) and (m/s) respectively in North, East and Down coordinate frame.\\ 
                               \hline
        \multirow{4}{*}{\parbox{1.2cm}{Sensors ($y_t$)}}
                                & IMU and Magnetometer & The two are used together to measure $\phi$ and $\omega$ \\ \cline{2-3}
                                & GNSS   & Only measures $p$\\\hline
        \multirow{3}{*}{\parbox{1.2cm}{Imp. Specific Argument ($\theta$)}}
                                & Controls & Actuation command to each rotors.\\\cline{2-3}
                                & Time Interval & Period between each EKF call in (ms).\\\cline{2-3}
                                & Error Covariance & Uncertainty of state values.\\\hline
        \end{tabular}
\end{table}

\psection{Result}
We found that using linear deep models worked the best for the surrogate as well as the spoofer.
The likely cause is due to SAVIOR's EKF being less complex than PX4.
We repeated the experiment 10 times and an example of attack performance of one experiment is plotted in \fig{\ref{fig:savior}}.
The plots shows that the spoofer achieved up to 2.5m more estimation error than the highest nominal estimation error when remaining below the detection threshold.
While the available research implementation of SAVIOR exhibited numerical instability, \papername{}'s ability to identify stealthy adversarial inputs within the narrow operational window before the instability manifests demonstrates resilience under the kind of constrained conditions a real world attacker would face.

\section{Discussion}
\label{sec:discussion}
We discuss the conditions required for \papername{} to work (\ie learnability of the target function), potential defenses against \papername{} and limitation of our methods.

\subsection{\ref{rq:stealthy}: What variables affect an attack's ``stealthiness''?}
\label{sec:rq:stealthy}
Our results reveal that the intrinsic factors (\eg sensor noise and imperfect vehicular dynamics modeling) are sufficient for \papername{} to evade detection, even under controlled laboratory conditions (\S\ref{sec:rq:realworld}).
Other variables (\eg environment) that introduce uncertainty in fact make stealth easier, as operators relying on reported positions cannot distinguish wind induced drift from attack induced deviation.
This disconnect between the UAV's local environment and the operator's remote perspective creates an exploitable gap.

Hence, \textit{building an anomaly detector tied to the vehicle's dynamics alone is likely insufficient to detect} \papername{}.
Instead, system-level protection of important variables~\cite{gruss2017kaslr,yoon2016taskshuffler}, detection of attack vectors~\cite{ismail2022tightly, xiang2024boosting, budak2025lightweight} and redundancy architectures~\cite{restart_abdi2018guaranteed, restart_kashinath2025groundhog} may be more effective.

\subsection{\ref{rq:meaningful}: What factors make attacks to be "meaningful"?}
\label{sec:rq:meaningful}
Missions impose space and time constraints;
space constraints limit the allowable deviation from the planned trajectory, while the time constraints specify mission deadline requirements.
We observe that attacks that violate these constraints are what make them ``meaningful''.

However, the strictness of those constraints vary depending on the mission type.
Delivery missions can tolerate path deviations as long as it can reach the destination in time (\ie taking a different path from point A to point B).
In contrast, safety critical missions, such as surveillance or search and rescue,  often have strict time and space constraints
(\eg scanning over a mission area within a specified time to detect humans), especially when coordination between multiple vehicles is required (\eg swarms).

Therefore, an attack capable of causing the most space deviation within the time constraint is considered to be the most ``meaningful''.
Our missions assume strong space constraints --- that the path taken by the vehicle is important, \eg surveillance or reconnaissance missions.
Although we didn't enforce a time constraint in our missions, our approach also demonstrated its capability of violating the time constraints as it achieved the highest maximum deviation in the shortest time (\S\ref{sec:rq:success}).

\subsection{Defending against \papername{}}
Because we exploit errors that naturally exist in vehicular systems, defending against \papername{} is very tricky when input and outputs can be collected.
One could try to prevent the payload from reaching the target system, \ie making it difficult to `learn' the target function with surrogate models or increasing sensor redundancy where multiple sensors provide similar overlapping information.
Another potential solution is via randomization \cite{yoon2016taskshuffler, chen2021indistinguishability} since it makes the mapping between the input and the output seem more disjoint.
Though in autonomous vehicle domains, applying randomization to state estimation may be difficult since the vehicular motion is bounded by standard models of physics.
{Overall, we highlight a fundamental tension \viz the determinism required for robust modeling, safety and certification is precisely what enables learnability, \ie makes our attack possible.}

\subsection{Limitations}
{We deliberately scope our evaluation to single-sensor spoofing to isolate and demonstrate the core vulnerability; multi-sensor spoofing introduces additional challenges such as maintaining consistency across sensors with different update rates, which we leave for future work.}
It is important to note that we can cause significant problems by spoofing just one sensor.
In the threat model, we assume that the adversary is aware of the state of the module.
We acknowledge that it is difficult to map every variable.
However, it can be alleviated in some cases: identifying regions of memory corresponding to module parameters can be achieved by setting unique parameter values with MAVLink \cite{mavlink} commands and searching for them.
Regions of memory for sensors can be discovered by manipulating sensor messages that reach the estimation module.
We {\em do not} consider stealthiness under post-flight analysis where a full detailed log of the vehicle is used to diagnose any potential issues with the vehicle.
The attack capability is limited when there is a drastic distribution shift between the data collection and deployment.
However, our evaluations demonstrate that \papername{} generalizes across environmental (wind \S\ref{sec:rq:weather}) and kinematic shifts (\eg random walk vs. circular missions), demonstrating resilience to distribution mismatches.

\section{Related Work}
\label{sec:related}

We now explore some relevant contemporary strategies for creating Adversarial Examples (AEs) and their deployment in autonomous systems, focusing especially on stealthy spoofing attacks and note how \papername{} differs from comparable work.

\psection{Spoofing Attacks}
Recent studies\cite{Fu2022Ad2Attack, urbina2016limiting, Sobh2021Adversarial, shapeshifter, slap, song2018physical} involving adaptive adversarial attack approaches \etc highlight the inherent brittleness of Deep Neural Networks~(DNNs) and consequently the susceptibility of anomaly detectors in various autonomous systems including UAV Autopilot systems.
The adversarial examples produced by \papername{} are adaptable and broadly applicable across a variety of systems since they do not depend on specific system states or sensors.

\psection{Detection and Defense Against Spoofing Attacks}
More recent anomaly detectors, such as SAVIOR~\cite{quinonez2020savior}, use EKF-based state estimation to detect anomalies using the residual.
The essence of their EKF is similar to that used in PX4.
Hence, we can apply our methods to generate another spoofer for this EKF and tune the loss functions to evade detection by such methods as demonstrated in \S\ref{sec:eval:savior}.
Most of the current methods to detect~\cite{hickling2023robust, kim2020security, kim2024systematic} and defend against spoofing attacks such as sensor watermarking\cite{mo2009secure, naha2023quickest} \etc work best with fixed spoofing strategies (\ie baseline attacks).
\papername{} makes such defenses moot since our spoof generation is a function of the actuation command and we demonstrated that a single model is capable of minimizing the residual under various kinematics.

\bibliographystyle{IEEEtran}
\bibliography{refs}

\appendix

\subsection{Referenced Missions}
\label{appendix:missions}
This section describes the procedures for missions referenced in \S\ref{sec:req:dc} and \S\ref{sec:eval:subsec:exp_param}: random walk, circle, linear and hold.

\psection{Random Walk}
It is used to capture as much diverse movement as possible to ensure that the collected data can be used to train the model meaningfully.
We get the setpoints for Brownian Motion by using a random walk algorithm on a 2D plane where the next setpoint is the current setpoint moved by a value sampled from 2D Gaussian with zero mean and unit variance as shown in \eq{\ref{eq:brownian}}:
\begin{equation}
    \label{eq:brownian}
    p_0 = (0, 0); p_{t+1} = p_t + \epsilon_{t+1}; \epsilon_{t+1} \sim \mathcal{N}((0,0), I)
\end{equation}

$p_0$ is the origin position with $0m$ east and $0m$ north respectively.
Identity matrix $I$ is the covariance matrix (\ie no covariance between east and north).
To add rotational kinematics into the dataset, we randomize the heading for each setpoint.
The UAV is considered converged to the setpoint if the vehicle is within $0.3m$ and $1^{\circ}$ of the setpoint.
Example trajectories for a random walk is on the website\footnote{\label{fn:paper_website}\paperwebsite{}}.

\psection{Circle}
Circle Mission simulates a surveillance mission where a UAV circles a designated area with a part of the UAV constantly facing toward the center.
Specifically, the UAV moves forward at $5m/s$ while rotating $30^{\circ}$ counter-clockwise resulting in a circle with $9.549m$ radius centered at $(-8.25m, 0.9m)$ east-north coordinate.
For real world, $0.5m/s$ forward with the same rotational velocity was used.

\psection{Linear}
Linear Mission represents a mission involving logistics such as delivery (\eg Amazon Air) where the UAV reaches each of the check points in a straight line.
Three setpoints are used to cover the movement solely along north (N) axis, east (E) axis and both axes.
At altitude of 10m, the UAV moves from the origin to (10m E, 0m N), then to (10m, 10m), and returning back to the origin (0m, 0m).
For real world, altitude of 0.5 with (1m, 0m), (1m,1m), then (0m,0m) was used.

\psection{Hold}
In all types of missions, the UAV may need to briefly hold its position.
The Hold Mission is accomplished by only sending a single setpoint: 0m north, 0m east of the starting point at 10m altitude.
Altitude was set to 0.5m for real world evaluation.

\subsection{Extended Kalman Filter Detail}
    \label{appendix:ekf}
    \label{sec:basm:ekf}
EKF has two stages: \textit{predict} and \textit{update}.
The \textit{predict} stage provides a meaningful state estimation based on the actuation command in between sensor updates.
The \textit{update} stage adjusts the state estimation based on sensors.

\psection{Predict}
This stage utilizes the previous actuation command to provide the latest state estimation.
For example, if the actuation command cause
a full rotation of the wheels of a car, then the distance that the vehicle travels should be equal to the circumference of the wheel.
The function that handles such calculations is the \textit{transition function}.

Formally, the function $f$ is the State Transition Function that predicts the State $\hat{x}_{t|t-1}$ at Time $t$, based on the Previous Estimation Update $\hat{x}_{t-1}$ and actuation command, as shown in \eq{\ref{eq:ekf_predict}}:

\noindent{}\begin{minipage}{\columnwidth}
\vspace{1.2em}
\begin{equation}
    \label{eq:ekf_predict}
    \eqnmarkbox[blue]{eq_ekf_pred_pred_state}{\hat{x}_{t|t-1}} =
    \eqnmarkbox[ForestGreen]{eq_ekf_pred_state_trans}{f}(
        \eqnmarkbox[NavyBlue]{eq_ekf_pred_previous_state}{\hat{x}_{t-1}},
        \eqnmarkbox[black]{eq_ekf_pred_actuation_command}{u_t})
\end{equation}
\annotate[yshift=0em]{left, below}{eq_ekf_pred_pred_state}{predicted state}
\annotate[yshift=0em]{right, below}{eq_ekf_pred_state_trans}{transition function}
\annotate[yshift=0.5em]{left, above}{eq_ekf_pred_previous_state}{previous state update}
\annotate[yshift=0.5em]{right, above}{eq_ekf_pred_actuation_command}{actuation command}
\end{minipage}

The formulation of $f$ differs from vehicle to vehicle based on the hardware(\eg motor) and the physical characteristics(\eg wheel size).
There can be an error in the modeling of the vehicle dynamics due to expensiveness of accurate modeling in vehicles with complex dynamics (\eg quadrotor drones) or due to noise in the system (\eg due to friction or environmental factors).

\psection{Update}
When new sensor data becomes available, the estimator calculates the \textit{expected} sensor value based on the predicted state using an observation function and compares the observed value with the expected value.
The disparity between the two is known as the \textit{residual}, and determines how much the predicted state
needs to adjust.
Additionally, the residual
is used to gauge the degree of abnormality in sensor measurements.

\noindent{}\begin{minipage}{\columnwidth}
    \vspace{2em}
    \begin{minipage}{\linewidth}
        \begin{equation}
            \label{eq:ekf_update_res}
            \eqnmarkbox[red]{n1}{r_t} = 
            \eqnmarkbox[NavyBlue]{n2}{y_t} - 
            \eqnmarkbox[Plum]{n3}{h}(
                \eqnmarkbox[Green]{n4}{\hat{x}_{t|t-1}}
            )
        \end{equation}
        \annotate[xshift=1em, yshift=0.5em]{left, above}{n1}{residual}
        \annotate[yshift=0.5em]{right, above}{n2}{sensor}
        \annotate[yshift=-0.25em]{left, below}{n3}{observation function}
        \annotate[yshift=-0.25em]{right, below}{n4}{predicted state}
        \vspace{1mm}
    \end{minipage}
    \begin{minipage}{\linewidth}
        \begin{equation}
            \label{eq:ekf_update_fuse}
            \eqnmarkbox[NavyBlue]{ekf_u1}{\hat{x}_t} =
            \hat{x}_{t|t-1} +
            K_t
            r_t
        \end{equation}
        \annotate[yshift=-0.25em]{right, below}{ekf_u1}{updated state}
        \vspace{1mm}
    \end{minipage}
\end{minipage}

Formally, when the sensor, $y_t$, is available, the \textit{residual} is calculated based on \eq{\ref{eq:ekf_update_res}} using the observation function $h$, which happens to be non-linear.
Therefore, linearization is performed to construct Kalman gain $K$ that is used to update the state, $\hat{x}$, as shown in \eq{\ref{eq:ekf_update_fuse}}.
$K$ adjusts how much influence the residual $r$ should have on the update.
EKF implementations often use $\chi^2$ anomaly detector as default (explained in \S\ref{sec:threatmodel:ad}).

\subsection{Implementation Detail}
\label{appendix:implementation}
\psection{Simulation Experiment Setup}
This subsection describes the implementation detail to run the experiments from \S\ref{sec:eval}.
We use PX4 (v1.13.1) Software-In-The-Loop (SITL) with Gazebo for physics simulation. 
The simulation uses default world with the default quadrotor model in PX4.
We use MAVSDK-python\cite{MAVSDK-Python} to send commands to the UAV.
We use a combination of TCP and mmap for exfiltration of the snapshots
from the PX4 to the attack server.
where TCP packets were used
to signal that data is ready for in the mmap.
The specification is elaborated in more detail on the website\footnoteref{fn:paper_website}.

\subsection{\papername{} Training Detail}
    \label{appendix:requiem_train}
    In \S\ref{sec:req:sm}, we used data augmentation for increase training set size.
    As an additional measure to prevent overfitting, we augmented data for each training iteration (\ie different $\eta$ for each training loop)
    for both surrogate training and spoofer training.
    The overall detail of the the training parameters for surrogates and the spoofer are shown in \tab{\ref{tab:surrogate_train_param}}.
    All three models used Adam \cite{kingma2014adam} for optimization with learning rate set to $10^{-4}$.

    \begin{table}[t]
        \centering
        \footnotesize
        \caption{\label{tab:surrogate_train_param}Surrogates and spoofer training parameters for PX4.
        }
            \begin{tabular}{|>{\raggedright}p{0.70in}||p{0.775in}|p{0.775in}|p{0.55in}|}\hline
        \textbf{Model Description}&
            Surrogate model for residual calculation &
            Surrogate model for velocity and position calculation respectively&
            Spoofer\\\hline
        
        \textbf{Input Size}&
            52 &
            54&
            52\\\hline

        \textbf{Input Description}&
            Table. \ref{tab:collected_data_type} &
            Table. \ref{tab:collected_data_type} with output from  prev surrogate&
            Table. \ref{tab:collected_data_type} \\\hline
        \textbf{Hidden Size} & 50 & 50 & 50\\\hline
        \textbf{\# of Hidden Layers}& 2 & 10& 5\\\hline
        \textbf{Activation Func}& ReLU & ReLU& ReLU\\\hline
        \textbf{Output Size}& 8 & 1 & 2\\\hline
        \textbf{Output Description}&
            Velocity and position residual and residual variance during the state update&
            State estimation of north velocity or north position resulting from the state update&
            Injection values for GNSS velocity and position \\\hline
        \textbf{\# of parameters}&
            8158 &
            28652&
            15502\\\hline
        \textbf{Epochs}& 50 & 50& 140\\\hline
        \textbf{Training Time (Hrs)}& 16& 24& 74 \\\hline 
        \textbf{\% time spent on Query}& 99.3 & 99.1 & 92.5 \\\hline 
    \end{tabular}

        \vspace{-2mm}
    \end{table}

    \psection{Spoofer} 
    Just to make sure that the surrogate output accurately during the spoofer's training, we use \alg{\ref{alg:spoofer_train}}.
    In lines \ref{alg:spoofer_train:roi}-\ref{alg:spoofer_train:roi:end}, the surrogate model is fined-tuned to ensure that its output (when used to train the spoofer) is close to the target function.

    Notice that the major bottleneck during training was the time spent on querying $F$ for surrogate training.
    Training the spoofer requires more queries since calculating deviation bias loss (\eq{\ref{eq:deviation_bias_loss}}) requires additional query of $F(s)$ on top of $F(s+G(s))$ for surrogate fine tuning. 
    To reduce the querying frequency, we make a heuristic assumption that the state estimator without adversarial influence would likely to estimate position close to the unspoofed GNSS position value.
    We apply the heuristic to \eq{\ref{eq:db_opt}} by replacing the  $F(s)$ term with $y^{GNSS}$ resulting in 
    
    \noindent{}\begin{minipage}[t][1.4cm][t]{\linewidth}
        \vspace{2mm}
        \begin{equation}
            \label{eq:modified_deviation_bias_loss}
            L_d=
            ReLU(
            \eqnmarkbox[Blue]{predicted}{P(y^{GNSS})} -
            \eqnmarkbox[Orange]{spoofed}{P(\hat{F}(s + G(s)))}
            )\\
        \end{equation}
        \annotate[yshift=-0.25em]{left, below}{predicted}{position reported by GNSS}
        \annotate[xshift=0.2em, yshift=0.5em]{left, above}{spoofed}{estimated position resulting from spoof}
        \vspace{2mm}
    \end{minipage}
    
    \noindent{}We used parameters $\delta=0.025$ (line \ref{alg:spoofer_train:roi}) for fine-tuning and $T=0.1$ for budget loss calculation.

   \begin{algorithm}[t]
        \caption{Spoofer Training}
        \label{alg:spoofer_train}
        \begin{algorithmic}[1]
    \begin{scriptsize}
    \FOR{$e \in Epoch$}
        \STATE \COMMENT{Enumerate over the collected data}
        \FOR{$s \in S$}
            \STATE $\tilde{s} \gets s + \eta$ \COMMENT{Dynamic data augmentation}
            \STATE \COMMENT{Ensure that the surrogate model is within $\delta$}
            \WHILE{$L_s > \delta$}\label{alg:spoofer_train:roi}
                \STATE $s_f \gets F(\tilde{s}+G(\tilde{s}))$
                \STATE $L_s \gets \mathcal{L}(s_q, \hat{F}(\tilde{s}+G(\tilde{s})))$
                \STATE $\texttt{Backprop}(L_s)$ \COMMENT{Update the surrogate model}
            \ENDWHILE\label{alg:spoofer_train:roi:end}
            \STATE \COMMENT{Train the spoofer}
            \STATE $l \gets \mathcal{L}_{G}(\hat{F}(\tilde{s}+G(\tilde{s})))$ \COMMENT{From \eq{\ref{eq:db_opt}}}\label{line:spoofer_train:spoofer_loss}
            \STATE $\texttt{Backprop}(l)$ \COMMENT{Updates the spoofer}
        \ENDFOR
    \ENDFOR
    \RETURN $G$
    \end{scriptsize}
\end{algorithmic}

    \end{algorithm}

\subsection{Total Bregman Divergence}
\label{appendix:tbd}
Total Bregman Divergence (\TBD{}) is used to quantify the difference between two different shapes~\cite{liu2010total}.
As mentioned in \S\ref{sec:eval:subsec:metrics}, it was used to to show that a deviated trajectory due to an attack is significant by comparing the true trajectory against the nominal.
By treating trajectories as shapes, we can quantify the difference between two trajectories.

The difference is calculated by representing the trajectory as a Gaussian mixture model, $p$, with a set of $n$ Gaussian distributions and comparing the models between the two trajectories.%
Each Gaussian ($\mathcal{N}$) has mean $\mu_i$, co-variance $\Sigma_i$ and weight $a_i$.
The \TBD{} between two mixture models,$p_k$ and $p_h$, is defined as \eq{\ref{eq:tsl:d}}.

\noindent{}
\begin{minipage}{\columnwidth}
    \vspace{-2mm}
    \begin{align}
        \label{eq:tsl:d}
        tBD(k,h) = d_{k,h} = \sum_i^na_i^{k}a_i^{h}\mathcal{N}(0,\mu_i^k-\mu_i^h,\Sigma_i^k+\Sigma_i^h)
    \end{align}
\end{minipage}
\noindent{}Therefore, to get the difference between two shapes (\ie total squared loss (tSL) of the \TBD{}) is calculated as shown in \eq{\ref{eq:tsl}}

\noindent{}
\begin{minipage}{\columnwidth}
    \vspace{-2mm}
    \begin{align}
        \label{eq:tsl}
        tSL(p_r, p_q) = \frac{d_{r,r}+d_{q,q}-2d_{r,q}}{\sqrt{q+4d_{q,q}}}
    \end{align}
\end{minipage}
Where $p_r$ is the reference mixture model (\ie nominal trajectory) and the $p_q$ is the query (\eg true trajectory).
Furthermore, we calculate the tSL for the reported position to show that the observed trajectory is similar to the mission trajectory.

\subsection{Effects of Weather}
\label{app:ref_scenario}

\begin{figure}
    \centering
        \begin{minipage}{\columnwidth}
            \centering
            \def\legendPadding{\hspace{2.5mm}}
            \footnotesize
            \begin{tabular}{|@{\legendPadding}l@{\legendPadding}l@{\legendPadding}l@{\legendPadding}|}\hline
                \begin{tikzpicture}[scale=1]
                    \draw[line width=0.5mm, color=black] (0,0) -- (0.5,0);
                \end{tikzpicture}
                : Planned 
                &
                \begin{tikzpicture}[scale=1]
                    \draw[line width=0.5mm, normal] (0,0) -- (0.5,0);
                \end{tikzpicture}
                : Nominal
                &
                \begin{tikzpicture}[scale=1]
                    \draw[dashed, line width=0.5mm, color=op ] (0,0) -- (0.5,0);
                \end{tikzpicture}
                : System POV
                \\
                \begin{tikzpicture}[scale=1]
                        \draw[line width=0.5mm, attack] (0,0) -- (0.5,0);
                        \end{tikzpicture}
                        : Stealthy
                &
                    \begin{tikzpicture}
                        \path (0,0) pic[line width=0.3mm, rotate = 0] {cross=3pt};
                    \end{tikzpicture}: Stealth Loss Point
                &
                \begin{tikzpicture} [scale=1]
                            \draw[dashed, line width=0.5mm, attack] (0,0) -- (0.5,0);
                        \end{tikzpicture}
                        : Overt
                \\\hline
            \end{tabular}
        \end{minipage}
    \begin{minipage}{0.49\columnwidth}
        \centering
        \begin{subfigure}{\linewidth}
            \includegraphics[height=3.5cm, width=\linewidth]{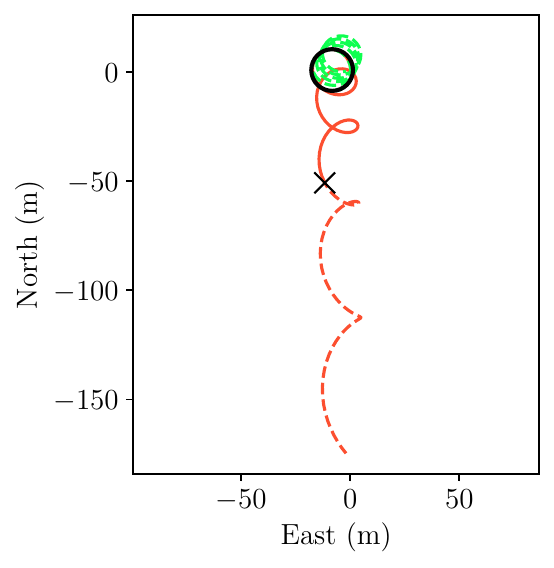}
            \caption{\textit{DB [NORTH WIND]}}
            \label{fig:db_windy:north}
        \end{subfigure}
    \end{minipage}
    \begin{minipage}{0.49\columnwidth}
        \centering
        \begin{subfigure}{\linewidth}
            \includegraphics[height=3.5cm, width=\linewidth]{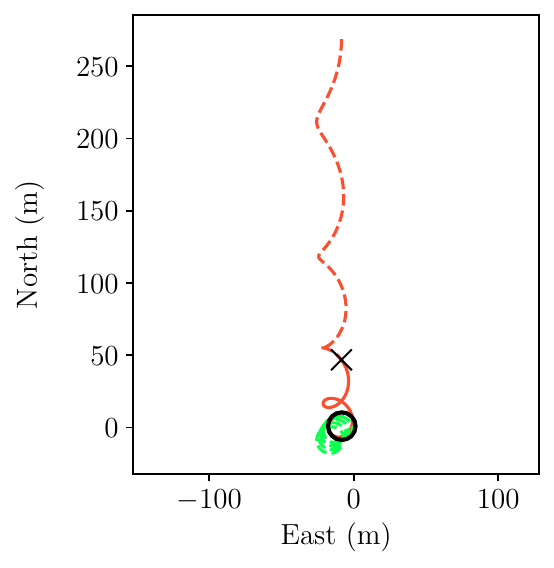}
            \caption{
            \textit{DB [SOUTH WIND]}
            }
            \label{fig:db_windy:south}
        \end{subfigure}
    \end{minipage}
    \caption{
            Comparison of wind direction affecting the DB attack shows that direction of deviation is opposite to the wind in both cases (\subref{fig:db_windy:north}) and (\subref{fig:db_windy:south}) but achieves greater deviation under southern wind.
            [NORTH WIND] On average over 10 trials, DB achieves 48.63m of stealthy deviation and 173.13m of overt deviation.
            [SOUTH WIND] On average, DB achieves 38.73m of stealthy deviation and 286.84m of overt deviation.
    }
    \label{fig:db_windy}
\end{figure}

\begin{figure}
    \centering
        \begin{minipage}{\columnwidth}
            \centering
            \def\legendPadding{\hspace{2.5mm}}
            \footnotesize
            \begin{tabular}{|@{\legendPadding}l@{\legendPadding}l@{\legendPadding}l@{\legendPadding}|}\hline
                \begin{tikzpicture}[scale=1]
                    \draw[line width=0.5mm, color=black] (0,0) -- (0.5,0);
                \end{tikzpicture}
                : Planned 
                &
                \begin{tikzpicture}[scale=1]
                    \draw[line width=0.5mm, normal] (0,0) -- (0.5,0);
                \end{tikzpicture}
                : Nominal
                &
                \begin{tikzpicture}[scale=1]
                    \draw[dashed, line width=0.5mm, color=op ] (0,0) -- (0.5,0);
                \end{tikzpicture}
                : System POV
                \\
                \begin{tikzpicture}[scale=1]
                        \draw[line width=0.5mm, attack] (0,0) -- (0.5,0);
                        \end{tikzpicture}
                        : Stealthy
                &
                    \begin{tikzpicture}
                        \path (0,0) pic[line width=0.3mm, rotate = 0] {cross=3pt};
                    \end{tikzpicture}: Stealth Loss Point
                &
                \begin{tikzpicture} [scale=1]
                            \draw[dashed, line width=0.5mm, attack] (0,0) -- (0.5,0);
                        \end{tikzpicture}
                        : Overt
                \\\hline
            \end{tabular}
        \end{minipage}
    \begin{minipage}{0.49\columnwidth}
        \centering
        \begin{subfigure}{\linewidth}
            \includegraphics[height=3.5cm, width=\linewidth]{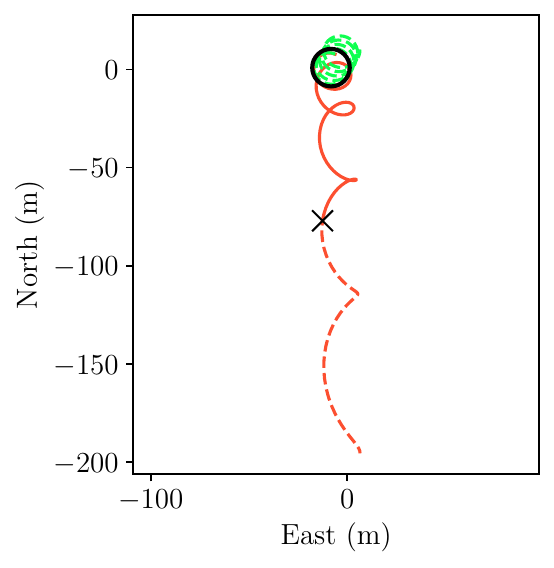}
            \caption{\textit{NC [NORTH WIND]}}
            \label{fig:nc_windy:north}
        \end{subfigure}
    \end{minipage}
    \begin{minipage}{0.49\columnwidth}
        \centering
        \begin{subfigure}{\linewidth}
            \includegraphics[height=3.5cm, width=\linewidth]{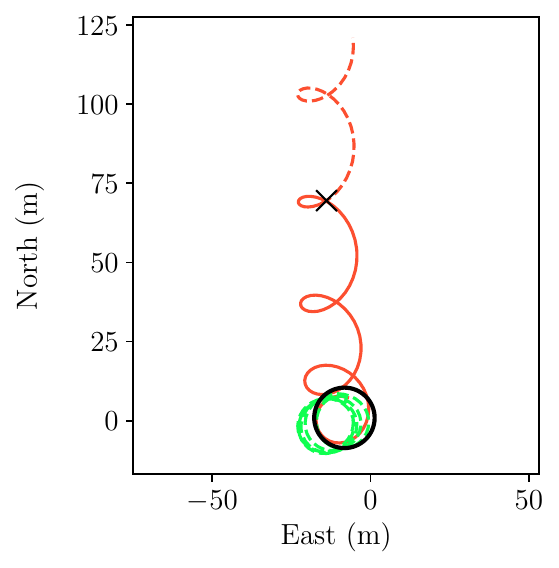}
            \caption{
            \textit{NC [SOUTH WIND]}
            }
            \label{fig:nc_windy:south}
        \end{subfigure}
    \end{minipage}
    \caption{
            Wind also makes NC attack to cause deviation opposite to the wind as shown in 
            (\subref{fig:nc_windy:north})  and (\subref{fig:nc_windy:south}).
            [NORTH WIND] On average over 10 trials, NC achieves 82.44m of stealthy deviation and 183.67m of overt deviation.
            [SOUTH WIND] On average, NC under southern wind achieves 73.85m of stealthy deviation and 140.04m of overt deviation.
    }
    \label{fig:nc_windy}
\end{figure}

This section continues the discussion from \S\ref{sec:rq:weather}.
The NC attack's prevention of the sensor from correcting the state estimation results in the control system overcompensation, thus causing deviations as shown in \fig{\ref{fig:nc_windy}} of \app{\ref{app:ref_scenario}}.
Comparing among the DB and NC attacks, the latter fared better since it remains stealthy for a longer time than DB.
However, DB achieves a larger deviation, at the cost of stealthiness: we can see the effect of the DB's bias effort in \fig{\ref{fig:db_windy:north}} where the UAV deviates less towards south and \fig{\ref{fig:db_windy:south}} where it deviates more towards north.
Both methods of \papername{} {\em still resulted in a significant stealthy deviation prior to becoming overt}.

\begin{figure}
    \centering
    
    \begin{minipage}{\columnwidth}
            \centering
            \def\legendPadding{\hspace{2.5mm}}
            \footnotesize
            \begin{tabular}{|@{\legendPadding}l@{\legendPadding}l@{\legendPadding}l@{\legendPadding}l@{\legendPadding}|}\hline
                \begin{tikzpicture}[scale=1]
                    \draw[line width=0.5mm, color=black] (0,0) -- (0.5,0);
                \end{tikzpicture}
                : Planned 
                &
                \begin{tikzpicture}[scale=1]
                    \draw[line width=0.5mm, normal] (0,0) -- (0.5,0);
                \end{tikzpicture}
                : Nominal
                &
                \begin{tikzpicture}[scale=1]
                    \draw[dashed, line width=0.5mm, color=op ] (0,0) -- (0.5,0);
                \end{tikzpicture}
                : System POV
                &
                \begin{tikzpicture}[scale=1]
                        \draw[line width=0.5mm, attack] (0,0) -- (0.5,0);
                        \end{tikzpicture}
                        : Stealthy
                \\\hline
            \end{tabular}
        \end{minipage}

    \begin{minipage}{0.49\columnwidth}
        \centering
        \begin{subfigure}{\linewidth}
            \includegraphics[height=3.5cm, width=\linewidth]{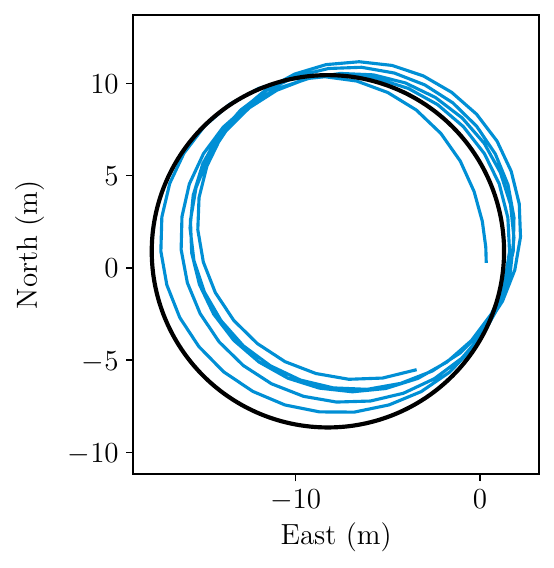}
            \caption{\textit{Nominal trajectory [NORTH WIND]}}
            \label{fig:windy_north:normal_traj}
        \end{subfigure}
    \end{minipage}
    \begin{minipage}{0.49\columnwidth}
        \centering
        \begin{subfigure}{\linewidth}
            \includegraphics[height=3.5cm, width=\linewidth]{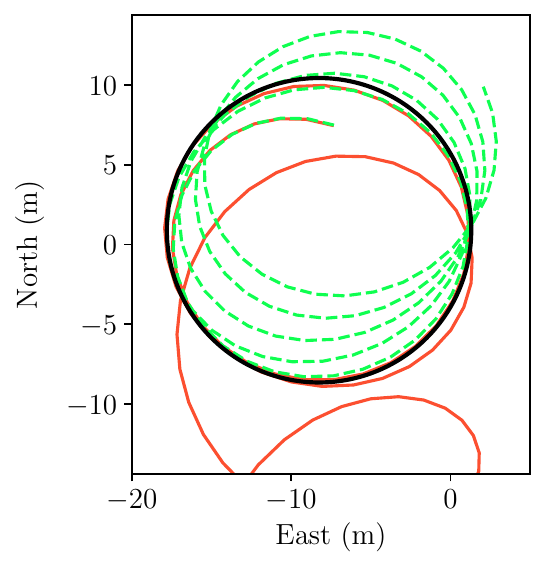}
            \caption{
            \textit{Zoomed trajectory from \fig{\ref{fig:nc_perf:sub:traj}} [NO WIND]}
            }
            \label{fig:windy_north:nc_zoom}
        \end{subfigure}
    \end{minipage}
    \caption{
        The deviations induced by \papername{} could be mistaken for weather effects, \eg wind.
        The nominal trajectory (blue) from a system with wind (\subref{fig:windy_north:normal_traj}) appears similar to System POV (green) without wind in (\subref{fig:windy_north:nc_zoom}).
        The operator in (\subref{fig:windy_north:nc_zoom}) cannot easily distinguish whether the deviation is due to weather or an attack.
    }
    \label{fig:windy_north_comp}
\end{figure}

\end{document}